\documentclass{article}

\usepackage{PRIMEarxiv}

\usepackage[utf8]{inputenc} 
\usepackage[T1]{fontenc}    
\usepackage{hyperref}       
\usepackage{url}            
\usepackage{booktabs}       
\usepackage{amsfonts}       
\usepackage{nicefrac}       
\usepackage{microtype}      
\usepackage{lipsum}
\usepackage{fancyhdr}       
\usepackage{graphicx}       
\usepackage{amsmath}
\usepackage{subfigure}
\usepackage{graphicx}
\graphicspath{{media/}}     
\usepackage{multirow}

\graphicspath{{Figures/}} 
\newcommand\Tau{\mathcal{T}}

\title{Optimizing a Transformer-based network for a deep learning seismic processing workflow
}

\author{
  Randy Harsuko and Tariq Alkhalifah \\
  King Abdullah University of Science and Technology \\
  Thuwal, Kingdom of Saudi Arabia \\
  \texttt{mochammad.randycaesario@kaust.edu.sa} \\
}

\begin{document}
\maketitle

\begin{abstract}
   StorSeismic is a recently introduced model based on the Transformer to adapt to various seismic processing tasks through its pretraining and fine-tuning training strategy. In the original implementation, StorSeismic utilized a sinusoidal positional encoding and a conventional self-attention mechanism, both borrowed from the natural language processing (NLP) applications. For seismic processing they admitted good results, but also hinted to limitations in efficiency and expressiveness. We propose modifications to these two key components, by utilizing relative positional encoding and low-rank attention matrices as replacements to the vanilla ones. The proposed changes are tested on processing tasks applied to a realistic Marmousi and offshore field data as a sequential strategy, starting from denoising, direct arrival removal, multiple attenuation, and finally root-mean-squared velocity ($V_{RMS}$) prediction for normal moveout (NMO) correction. We observe faster pretraining and competitive results on the fine-tuning tasks and, additionally, fewer parameters to train compared to the vanilla model.
\end{abstract}


\section{Introduction}
\label{sec:introduction}
While deep learning (DL)-based seismic processing approaches have been heavily proposed due to the power of DL in dealing with complex tasks and data, we have been overwhelmed by the maze of different architectures and techniques handling the various processing tasks. Recently, a trend in addressing many processing tasks with a single network has emerged, as various tasks help better constrain the training of the network. \cite{durall2022deep} demonstrated the ability of a single neural network (NN) in handling multiple processing tasks with separate training for each of the tasks. In fact, multi-task approaches have flourished as these networks tend to learn better with more objectives involved \cite{ruder2017overview}, with an example in the seismic sphere presented in \cite{ovcharenko2022multi}. Utilizing a Transformer network (popular in natural language processing applications) where traces are treated like words and shot gathers as sentences, \cite{harsuko2022storseismic} introduced a new framework for seismic processing called StorSeismic, in which the time axis is handled by learned 1D transformations and the relation between traces are determined by an attention mechanism \cite{vaswani2017attention}. They, specifically, demonstrated the potential of this network in transferring the knowledge of a pretrained model to perform various seismic processing tasks. However, the tasks were not connected to a proper seismic processing workflow. 

StorSeismic is mainly inspired by Bidirectional Encoder Representations from Transformers (BERT) \cite{devlin2018bert}. While the vanilla Transformer encoder was used (like in BERT), the embedding, where the input is transformed to hidden space, was adjusted to handle the input of shot gathers instead of sentences, and project seismic traces, instead of words, into the hidden space, followed by a sinusoidal positional encoding (PE) to inject positional information of the traces (offset). At the other end, the final layer (i.e., the prediction head) transforms back the hidden space to the original domain of the time samples. In spite of the promising results obtained on real data, the vanilla encoder and the fixed sinusoidal PE were not developed with seismic traces in mind. First, the computational cost of the dot-product attention within the encoder grows quadratically with sequence length \cite{tay2022efficient}, which is a well-known problem in the Natural Language Processing (NLP) community where the idea originated from, and it will be an issue as we handle large data. In the seismic community, the sequence length corresponds to the acquisition length (number of traces/channels), which in modern seismic acquisition, such as distributed acoustic sensing (DAS)-acquired data, could reach the order of thousands \cite{taweesintananon2021distributed}. Second, the sinusoidal PE brought by the original Transformer architecture \cite{vaswani2017attention} is shown to generalize poorly on longer sequences on NLP tasks \cite{press2021train}. With the recent advances in the Transformer architecture, various alternatives to the components within a Transformer model were proposed, potentially better suited for seismic data and could optimize the StorSeismic's performance. To tackle the quadratic complexity curse of Transformers, for example, diverse studies introducing different approaches were conducted, such as using low-rank attention \cite{wang2020linformer}, sparsification of the attention matrices \cite{beltagy2020longformer}, compression training in the student-teacher scheme (distillation) \cite{zaheer2020big}, etc. In search for a more expressive PE, researchers have been proposing various relative PE (RPE) techniques (e.g., \cite{raffel2020exploring}) that offer more flexibility than the standard PE because the positional information can be inferred from the nearby tokens (traces in our case) and might be trainable as well.

\begin{figure}[!h] 
\centering
\subfigure[]{\includegraphics[height=10cm]{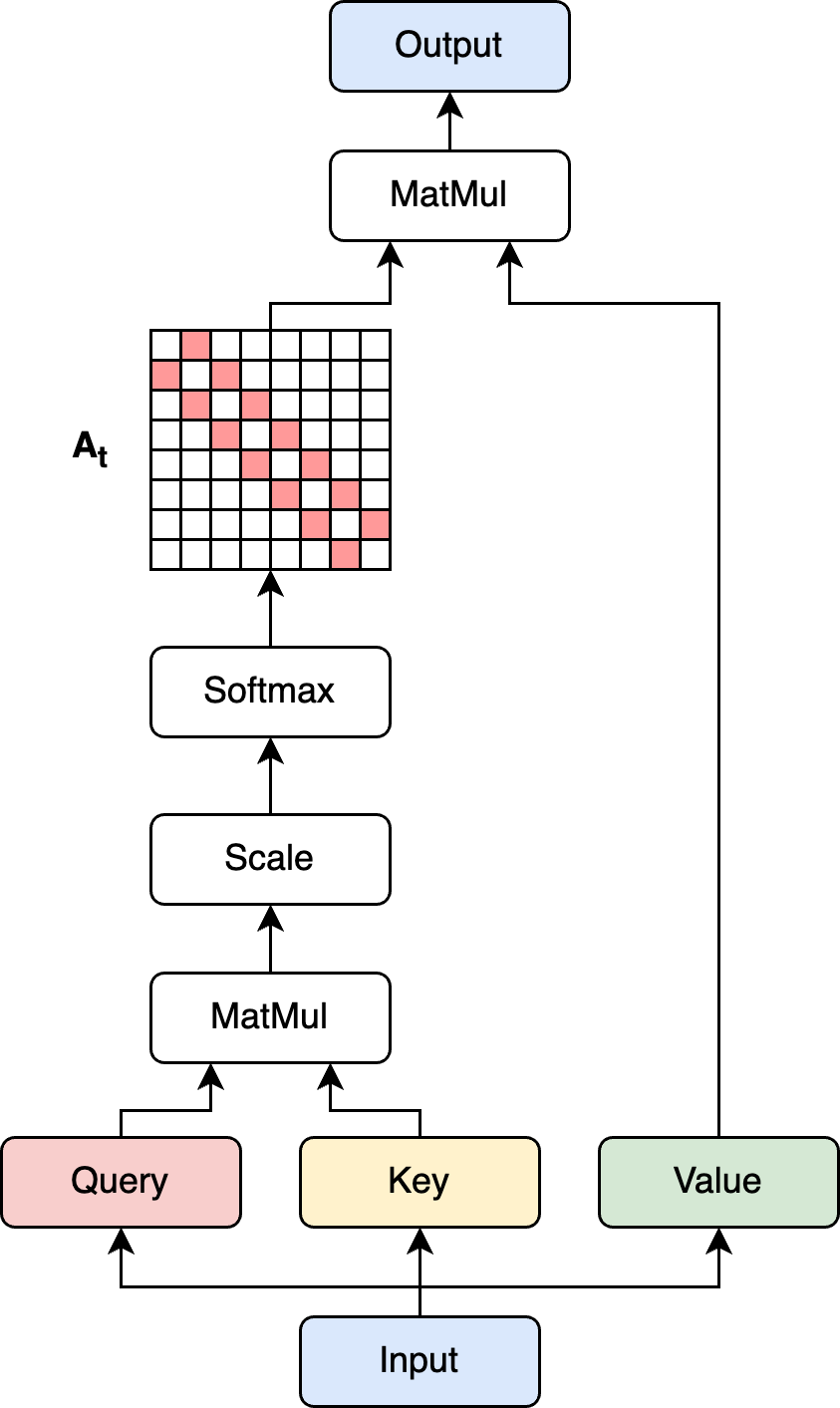} \label{fig:fig1a}}
\subfigure[]{\includegraphics[height=10cm]{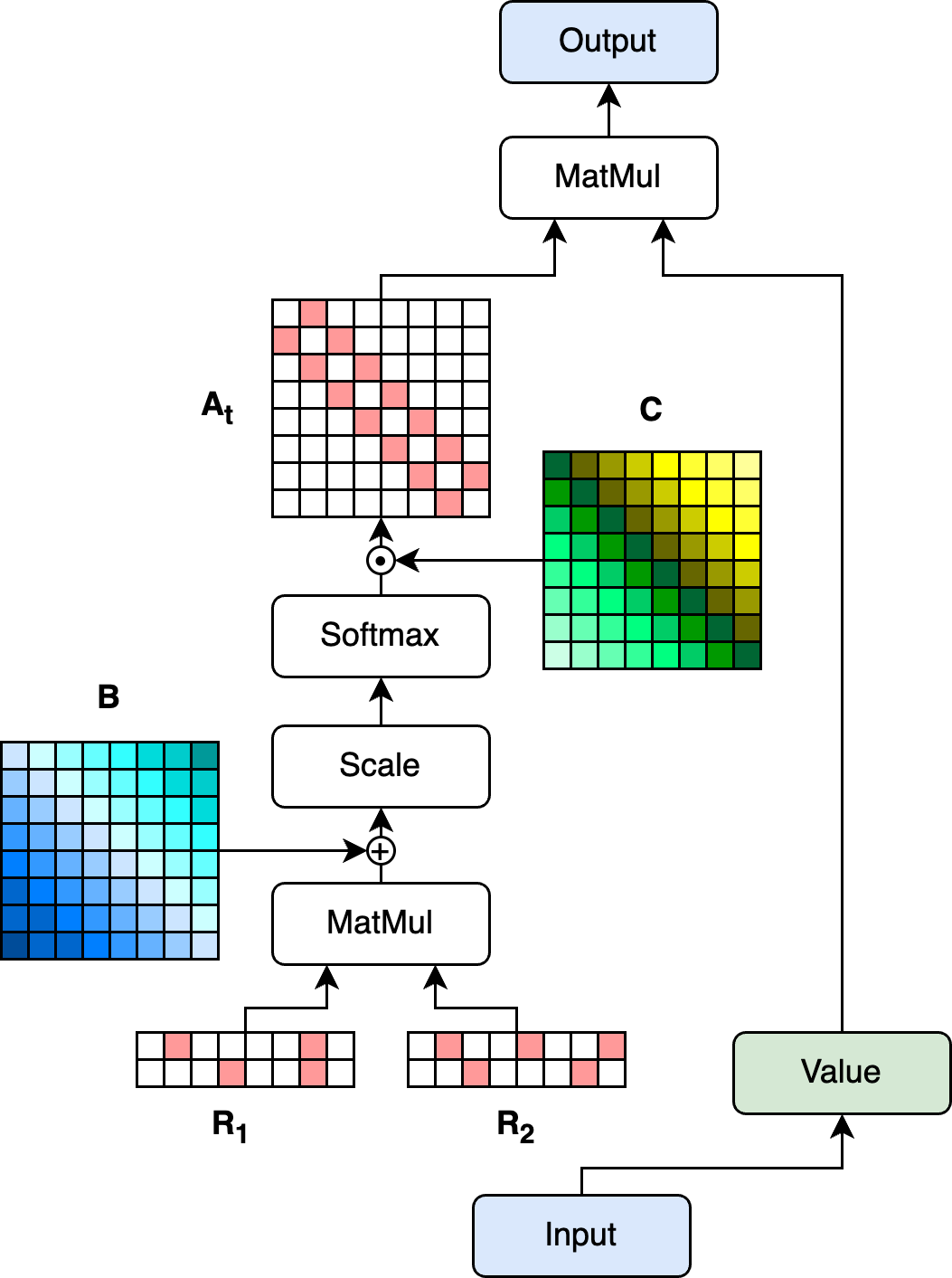} \label{fig:fig1b}}
\caption{(a) The vanilla scaled dot product attention and (b) the modified attention mechanism with factorized synthesizer ($R_1$, $R_2$), ALiBi ($B$), and URPE ($C$). The ($+$) symbol denotes matrix addition and the ($\cdot$) symbol denotes matrix element-wise product, while "MatMul" stands for matrix multiplication.} 
\label{fig:1} 
\end{figure}

Based on these motivations, we initiated a study on improving the architecture of StorSeismic for seismic applications, including testing the proposed variation's potential in a seismic processing workflow. The modifications are focused on the two core elements of a Transformer architecture, namely the attention mechanism and the positional encoding, as depicted in Figure \ref{fig:1}. In this study, we integrate StorSeismic into a conventional marine seismic data processing workflow, with challenging real field data included as part of the test.

The contributions of this paper can be summarized as follows:
\begin{enumerate}
    \item Testing modifications to the vanilla StorSeismic model, which yields a more efficient and effective Transformer-based network for seismic processing tasks. These modifications include a learnable positional encoding and low-rank attention matrices as replacements for the vanilla ones.
    \item We demonstrate the implementation of the proposed network in a conventional seismic processing workflow, particularly for marine seismic data, from denoising to obtaining post-stack data.
    \item We extensively compare the proposed modifications with the vanilla architecture on realistic Marmousi and field data.
\end{enumerate}

This paper is organized as follows. In Section \ref{sec:theory}, we briefly review the architecture of StorSeismic with emphasis on its positional encoding and attention mechanism, then propose modifications to them. Next, in Section \ref{sec:results}, we test the proposed modifications on Marmousi and field data and compare them with the vanilla architecture. We perform ablation studies and present them in Section \ref{sec:discussions}. Finally, we conclude the study with remarks and suggestions in Section \ref{sec:conclusions}.

\section{Theoretical Framework}
\label{sec:theory}
\subsection{Model architecture}
\label{sec:theory_1}
\vspace{-2.5mm}
The vanilla StorSeismic model \cite{harsuko2022storseismic}, adopted from the Transformer's encoder \cite{vaswani2017attention}, is composed of three main parts: the embedding block, encoder block(s), and prediction head. The model size is determined by the number of encoder blocks ($L$), the hidden dimension size ($H$), and the number of attention heads ($A$) with shot gathers $d$ of size given by the number of traces/channels $X$ and time samples $\Tau$ as the inputs. We will focus on studying the embedding and the encoder blocks. In the embedding block, for the network to recognize the position of a trace, a sinusoidal PE \cite{vaswani2017attention} $E_2$ is added to the linearly projected data $E_1 = W_{E_1}d + B_{E_1}$:
\begin{equation}
    E_2(pos, i) =
    \begin{cases}
        \sin(pos/10000^{2i/H}), & \text{if $i$ is even} \\
        \cos(pos/10000^{2i/H}), & \text{if $i$ is odd}
    \end{cases}
\end{equation}
where $pos$ is the relative position of the traces (i.e., offset index). The self-attention operation in the attention block, which is the key operation in Transformers, is defined as $Y = A_t V$ where $V \in \mathbb{R}^{X \times H}$ is the linearly transformed input. The vanilla attention matrix $A_t$ is given by:
\begin{equation}
    A_t = \text{softmax}(QK^T / \sqrt{H}).
\end{equation}
where $Q, K \in \mathbb{R}^{X \times H}$ are all matrices consisting of the linear projections of the input (query and key, respectively, with the linear projections having their own learned weights). The softmax operator normalizes the weights so that the action of $A_t$ on $V$ constitutes a weighted summation of the traces (or their transformation), and $QK^T$ is a measure of similarity between the traces (or their key features). The vanilla self-attention mechanism is illustrated in Figure \ref{fig:fig1a}.

The training of the model includes two steps: pretraining and fine-tuning. In the pretraining, the seismic data are fed into the network as shot gathers, in which some of the traces are masked, whereas the labels correspond to the original unmasked shot gathers. Therefore, the task in the self-supervised training is a reconstruction task as the loss is measured at the masked traces' location, thus forcing the network to learn the relation between the traces within a shot gather.  Afterwards, the appropriate input-label pairs are used to fine-tune the pretrained model based on a chosen task (e.g., noisy-clean pairs for denoising). 

\subsection{Modified positional encoding and attention mechanism}
\label{sec:theory_2}
\vspace{-2.5mm}
A major drawback of the sinusoidal PE in the vanilla implementation is its fixed nature, not adapting to the data. An alternative is to employ a relative PE (RPE) where the positional information is inferred relative to the location of the queried traces. \cite{press2021train} proposed attention with linear biases (ALiBi), in which the PE in the embedding is removed and replaced by a "slope"-based RPE in every attention head. The slopes are defined by variables multiplied by a linear increase (or decrease) of the weight of the traces away from the queried trace. Here, we use the \textit{non-symmetrical, learnable} variant of the ALiBi, in which the slopes are asymmetric on the queried traces, and more importantly, the variables for the slopes are learned. In addition, \cite{luo2022your} proposed a positional matrix complementary to the RPE, called universal RPE (URPE), which has been proven more effective on NLP tasks. The ALiBi and URPE are incorporated in Equation 2 as follows:
\begin{equation}
    A_t = \text{softmax}((QK^T + B) / \sqrt{H}) \cdot C.
\end{equation}
where $B \in \mathbb{R}^{X \times X}$ is the ALiBi and $C \in \mathbb{R}^{X \times X}$ is the URPE, which is a learnable Toeplitz matrix. When ALiBi and/or URPE are used, $E_2$ (Equation 1) is set to a zero-matrix (no positional encoding).

The trace-to-trace interaction occurs in the $Q$ and $K$ multiplication inside the self-attention mechanism, which is one of the key features of the architecture. However, this operation is a potential bottleneck when the acquisition lines (number of channels) are large. We could bypass this operation by replacing it with learnable weights $W \in \mathbb{R}^{X \times X}$ called \textit{synthesizer} \cite{tay2021synthesizer}:
 \begin{equation}
    A_t = \text{softmax}((W + B) / \sqrt{H}) \cdot C.
\end{equation} 
In this work, we test the \textit{factorized random} variant of the synthesizer, which involves the multiplication of two rectangular matrices to form a low-rank matrix $W$ (i.e., $W = R_1 R_2^T \mid R_1, R_2 \in \mathbb{R}^{X \times k},\; k << X$). 
Two advantages of using this form of the synthesizer: 1) Reducing the computational complexity by bypassing the projection step of potentially huge matrices ($Q$ and $K$) and their multiplication, and; 2) Reducing the number of neural network parameters of each block from a factor of $2H^2$ to a factor of $2AXk$. These modifications to the vanilla architecture are illustrated in Figure \ref{fig:fig1b}.

\subsection{The sequential implementation}
\label{sec:theory_3}
\begin{figure}[!h]
    \centering
    \includegraphics[width=0.8\textwidth]{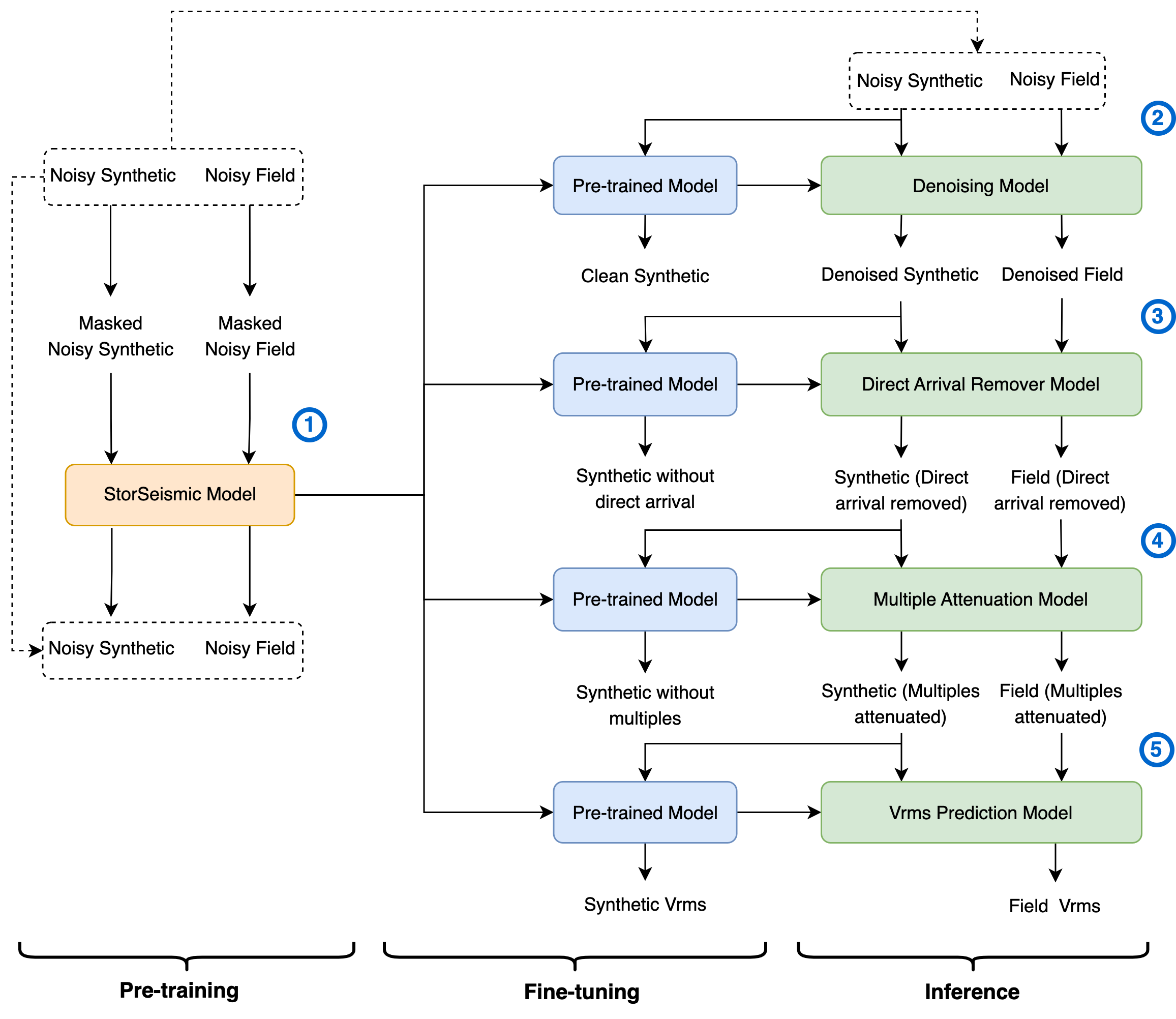}
    \caption{Schematic diagram of the fully deep learning-based seismic processing workflow.}
    \label{fig:fig2}
\end{figure}

Assuming that a post-stack seismic section is our end product, a typical processing workflow for marine seismic data may include denoising (e.g., with $f-k$ filters), removing direct arrival (e.g., with muting), attenuating surface multiples (e.g., with surface-related multiple elimination, SRME), and lastly estimating $V_{RMS}$ (e.g., with semblance velocity analysis) followed by normal moveout (NMO) correction and stacking. Here, we will replace the conventional methods with a single neural network, making the workflow entirely data-driven. This fully DL-based processing workflow is illustrated in Figure \ref{fig:fig2}, which is detailed as follows: 
\begin{enumerate}
    \item The pretraining dataset, composed of a mix of synthetic and field shot gathers, is partially masked trace-wise and used as the input to the network. The original, unmasked shot gathers are used as the corresponding labels in the pretraining. In other words, this pretraining procedure is self-supervised, and it is dedicated to feature-learning of both synthetic and label-less field data. 
    \item Our decided first task is to denoise the shot gathers. At this stage, we add realistic field noise to the synthetic data to form the input to the network and use the original synthetics (clean) as labels. We fine-tune the pretrained model to learn to denoise. The fine-tuned model is then used to denoise the synthetic and field shot gathers. 
    \item For imaging applications, the direct arrivals are often muted. Hence, we will remove them by fine-tuning the pretrained model. The input to this fine-tuning task is the denoised synthetic shot gathers from the output of step 2, and the labels are created by subtracting the clean synthetics from direct arrival-only data, which is obtained by modeling in a constant water velocity model. We then use the fine-tuned model to remove the direct arrivals on both the denoised synthetic and field data. 
    \item We next utilize the pretrained model from step 1 to attenuate the surface-related multiples, which, in conventional processing, are typically unwanted. The input to the fine-tuning is the synthetic shot gathers from the previous step (denoised and direct arrivals removed). We obtain the labels for this task by re-simulating the synthetic shots, but this time we replaced the free surface with an absorbing boundary condition, followed by direct arrival subtraction. After fine-tuning, we apply the model to attenuate the multiples in the data from the previous step (synthetic and field). 
    \item For stacking, we often apply velocity analysis to obtain the NMO (stacking) velocities needed in NMO correction. As an approximation, we will use the $V_{RMS}$ as labels \cite{fabien2020seismic}, given by 
    \begin{equation}
        V_{RMS \; N} = \sqrt{\frac{\Sigma_{i = 1}^N V_i^2 \Delta t_i}{\Sigma_{i = 1}^N \Delta t_i}}
    \end{equation}
    where $V_i$ is the velocity in the $i$-th layer, and $\Delta t_i$ is the two-way travel time (or here the time sampling interval) in that layer, for a model with $N$ time samples. Again, we use the pretrained model as a starting point in the fine-tuning to perform a direct mapping from the synthetic samples (shot gathers) from the previous step (denoised, direct arrivals removed, and multiples attenuated) to their corresponding $V_{RMS}$ profiles (averaged laterally over each of the shot positions). The final output of these processing steps is the $V_{RMS}$ profiles of the field data. 
\end{enumerate}

\section{Results}
\label{sec:results}
We test 5 NN models, each configured with $H = 512$, $L = 8$, $A = 8$: 1) Vanilla attention + sinusoidal PE; 2) Vanilla attention + ALiBi; 3) Vanilla attention + URPE; 4) Vanilla attention + ALiBi + URPE, and; 5) Synthesizer + ALiBi + URPE. For models with factorized synthesizer, we use $k = 16$ (refer to Section \ref{sec:discussions_2}). Table \ref{tab:tab1} summarizes the number of parameters for each model.

\begin{table}[!h]
    \centering
    ~\clap{
    \begin{tabular}{c|c|c|c|c|c|c}
        Model & Attention & $k$ & PE & RPE & \# of params. & \% params. \\
        \hline\hline
        Vanilla + PE (baseline) & vanilla & - & sinusoidal & - & 25,606,008 & 100.0 \\
        Vanilla + ALiBi & vanilla & - & - & ALiBi & 25,606,136 & 100.0\\
        Vanilla + URPE & vanilla & - & - & URPE & 25,647,480 & 100.2\\
        Vanilla + ALiBi + URPE & vanilla & - & - & ALiBi \& URPE & 25,647,608 & 100.2\\
        Syn. + ALiBi + URPE & synthesizer & 16 & - & ALiBi \& URPE & \textbf{22,108,664} & \textbf{86.3}
    \end{tabular}
    }
    \caption{The detailed configuration of each of the tested models and their corresponding number of parameters. The last column (\% params.) shows the number of parameters relative to the baseline model.}
    \label{tab:tab1}
\end{table}

The Syn. + ALiBi + URPE model has already gained one advantage at this stage. It possesses less parameters compared to the Vanilla + PE model (13.7\% less). This is attributed to the low-rank attention utilized in the Syn. + ALiBi + URPE model, as opposed to the vanilla full-rank attention. ALiBi and URPE introduced slightly more (yet comparably negligible) parameters to the model because the positional information is learned, as explained in Section \ref{sec:theory_2}.

\subsection{Marmousi example}
\label{sec:results_1}
As mentioned earlier, we utilize the well-known Marmousi model to test the method, and we consider the data simulated from the Marmousi model with added realistic noise as the "field" data. We use this model to represent the field data as the answers (labels) for the Marmousi are available, but used only for evaluating the accuracy of the approach. We simulate 450 shots with a spacing of 25 m in a streamer acquisition setup, where each shot gather contains 324 receivers spaced by a 25 m gap. The source wavelet is a Ricker wavelet with a peak frequency of 7 Hz. To create the synthetic data, first, we take a vertical velocity profile at location 10.5 km from the Marmousi and consider this as our well-log measurement. Then, we generate 2,048 2D random velocity models based only on the information from the well data. Acquisition setup and source characteristics are similar to the field (Marmousi) data, except that for each model, we only simulate three shots selected randomly, resulting in 6,144 synthetic shot gathers. We also add realistic noise sampled from the actual field data from offshore Australia (Section \ref{sec:results_2}) to the synthetic and field (Marmousi) data. Note that for this experiment, we used an acoustic simulation \cite{richardson_alan_2023} for testing purposes.

\begin{figure}[!h]
    \centering
    \includegraphics[width=0.65\textwidth]{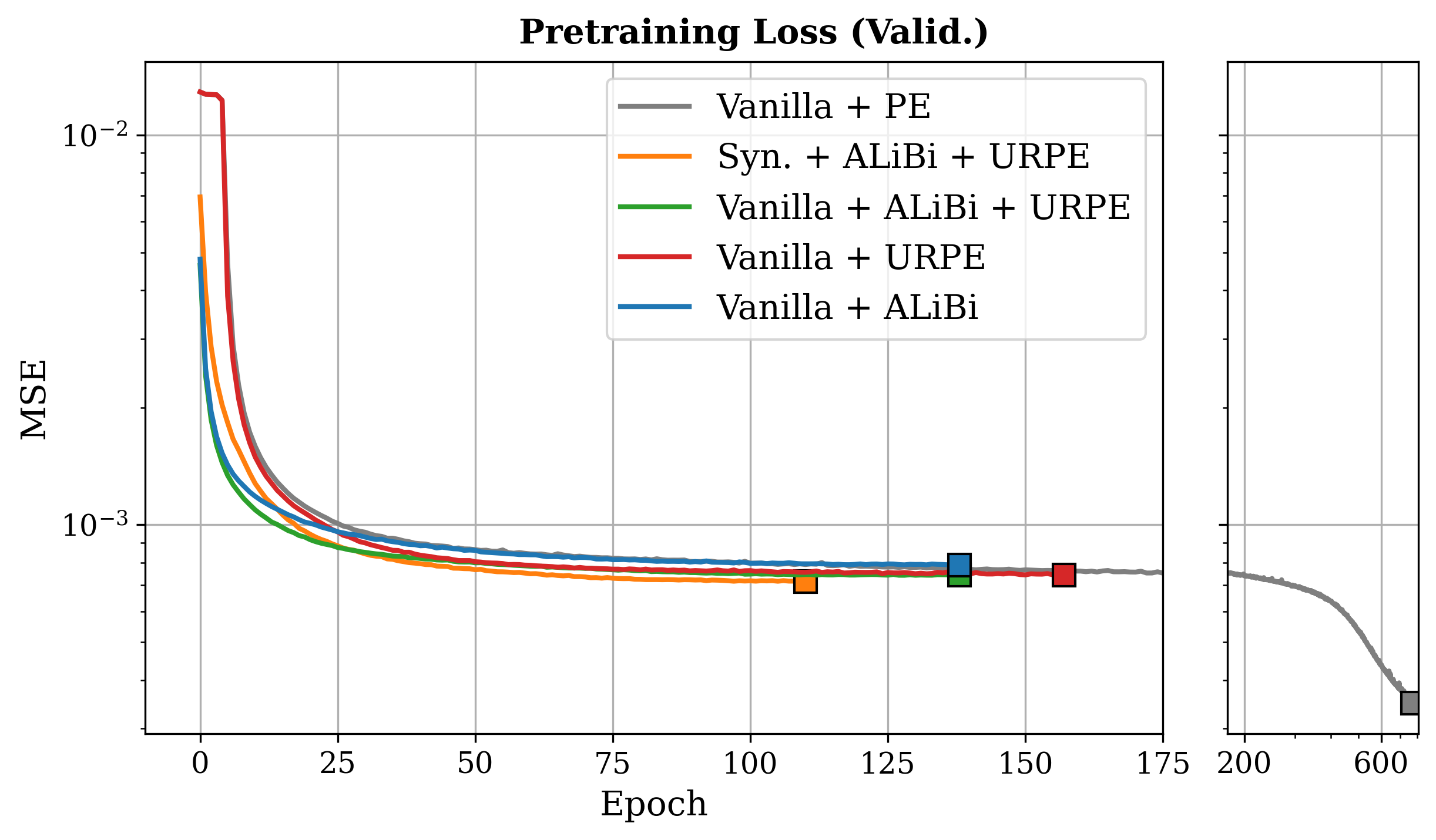}
    \caption{Pretraining loss curves of Vanilla + PE model (grey line), Syn. + ALiBi + URPE (orange line), Vanilla + ALiBi + URPE (green line), Vanilla + URPE (red line), and Vanilla + ALiBi (blue line) on validation set.}
    \label{fig:fig3}
\end{figure}

In the pretraining, we take a subset of the synthetic data to produce an equal mix of the synthetic and field data shot gathers \cite{harsuko2022storseismic}; then, we split them into 720 training and 180 validation samples. Afterwards, we perform data augmentation (shifting in time and reversing the polarity) and apply trace-wise masking with a 15\% masking probability, which expands our data to 43,200 training and 10,800 validation samples. We use the rectified Adam (RAdam) optimizer with a learning rate of $5\cdot10^{-4}$, a batch size of 128, and an L2 loss at the masked traces to pretrain each model on a single NVIDIA A100 GPU. An early stopping module, where the training is stopped if there is no improvement after 10 epochs, is used in the pretraining. The loss curves on the validation set (Figure \ref{fig:fig3}) show that all non-vanilla models converged in less than 200 epochs, while it took 770 epochs (amounts to 22.8 hours) for the Vanilla + PE model to converge. Moreover, within the first few epochs, a lower validation loss was achieved on all modified models (except the Vanilla + URPE model) compared to the Vanilla + PE model. Although the Vanilla + PE model has the lowest final validation loss, this does not determine the performance in the fine-tuning tasks, as we will see below.

\begin{table}[!h]
    \centering
    ~\clap{
    \begin{tabular}{c|c|c|c|c|c|c|c|c|c}
        \multirow{3}{*}{Model} & \multicolumn{2}{c|}{Denoising} & \multicolumn{2}{c|}{Dedirect} & \multicolumn{2}{c|}{Demultiple} & \multicolumn{2}{c|}{$V_{RMS}$ prediction} & \multirow{3}{*}{\shortstack[l]{Time \\ (hr $\downarrow$)}} \\
        & \multicolumn{2}{c|}{(PSNR $\uparrow$)} & \multicolumn{2}{c|}{(PSNR $\uparrow$)} & \multicolumn{2}{c|}{(PSNR $\uparrow$)} & \multicolumn{2}{c|}{(MAE $\downarrow$)}\\
        \cline{2-9}
         & Valid. & Test & Valid. & Test & Valid. & Test & Valid. & Test \\
        \hline\hline
        Vanilla + PE (baseline) & \textbf{42.79} & 41.04 & \textbf{42.39} & 40.07 & 43.61 & 32.48 & 79.69 & \textbf{49.17} & 32.1 \\
        Vanilla + ALiBi & 41.46 & 40.80 & 41.29 & 40.63 & 41.91 & 29.76 & 76.61 & 84.74 & 14.1 \\
        Vanilla + URPE & 42.01 & \textbf{41.17} & 42.23 & 41.02 & \textbf{44.10} & \textbf{33.56} & 76.22 & 65.19 & 22.3 \\
        Vanilla + ALiBi + URPE & 41.93 & 40.68 & 41.77 & 40.67 & 43.35 & 33.19 & 85.95 & 71.97  & 18.7 \\
        Syn. + ALiBi + URPE & 41.33 & 41.02 & 40.97 & \textbf{41.14} & 43.12 & 33.26 & \textbf{67.03} & 61.23 & \textbf{14.0}
    \end{tabular}
    }
    \caption{Measured metrics on all fine-tuning tasks for each model. The numbers shown are PSNR for denoising, dedirect, and demultiple, and MAE for $V_{RMS}$ prediction, averaged over the whole corresponding set. The rightmost column (Time) represents the total training time (pretraining and fine-tuning, in hours) spent for each model.}
    \label{tab:tab2}
\end{table}

\begin{figure}[!h]
    \centering
    \includegraphics[width=1\textwidth]{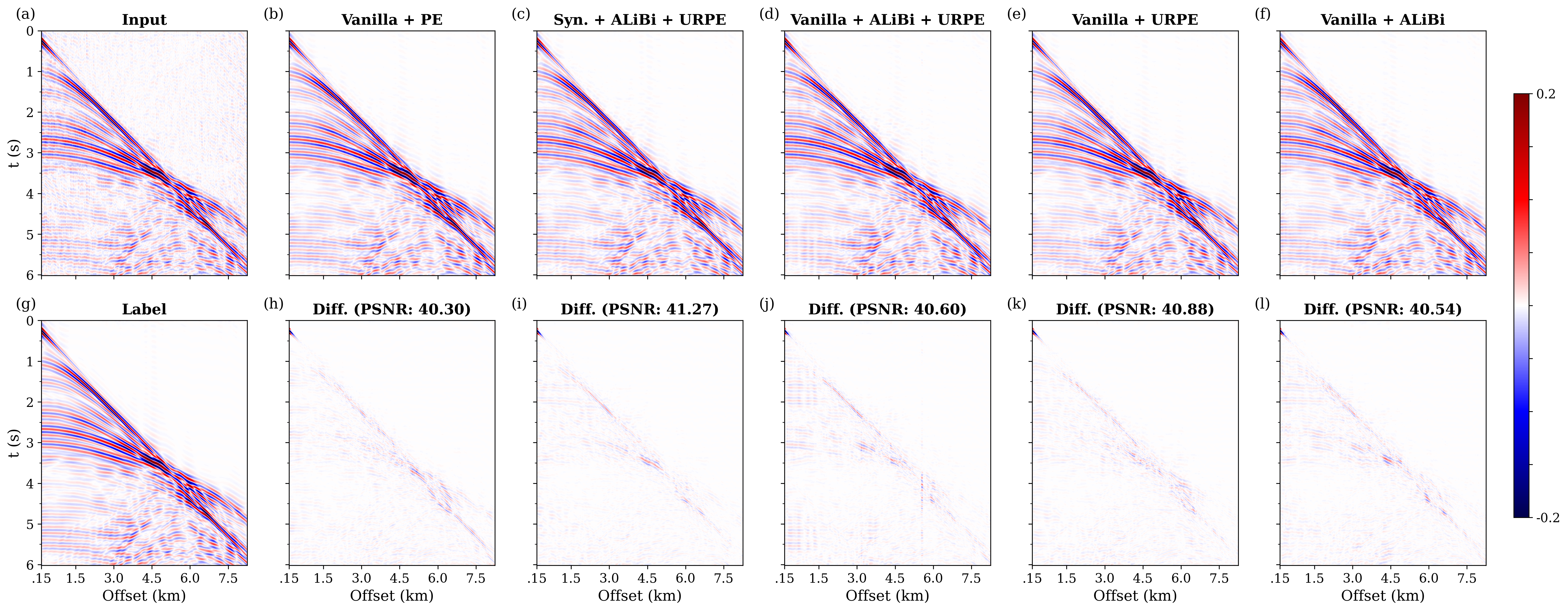}
    \caption{An example of denoising task on the Marmousi data. (a) Noisy input for each NN model and (g) its corresponding clean label. The output of: (b) Vanilla + PE; (c) Syn. + ALiBi + URPE; (d) Vanilla + ALiBi + URPE; (e) Vanilla + URPE, and; (f) Vanilla + ALiBi. (h--l) The difference between (b--f) and (g), respectively.}
    \label{fig:fig4}
\end{figure}

For all of the fine-tuning steps, we split the whole 6,144 synthetic shot gathers into 4,915 training and 1,229 validation samples. We only apply polarity reversal as an augmentation, which expands the data to 9,830 training and 2,458 validation samples. The field (Marmousi) data are considered as the test set. Unless otherwise stated, we use the RAdam optimizer (learning rate = $10^{-3}$), a batch size of 128, and L2 loss from this point forward. Additionally, we utilize the same early stopping module as used in the pretraining. To produce a denoising model, we fine-tune each of the pretrained models in which the loss is evaluated over the whole shot gather. The overall results for all models are listed in Table \ref{tab:tab2}. Though the Vanilla + PE model excels in the validation set, the Vanilla + URPE model achieved the highest PSNR (Peak Signal-to-Noise Ratio) in the test set. The Syn. + ALiBi + URPE model achieved similar performance with the Vanilla + PE model that comes in second. From sample results shown in Figure \ref{fig:fig4}, we could observe that the Syn. + ALiBi + URPE denoised data have less signal leakage than that of the Vanilla + PE model.

\begin{figure}[!h]
    \centering
    \includegraphics[width=1\textwidth]{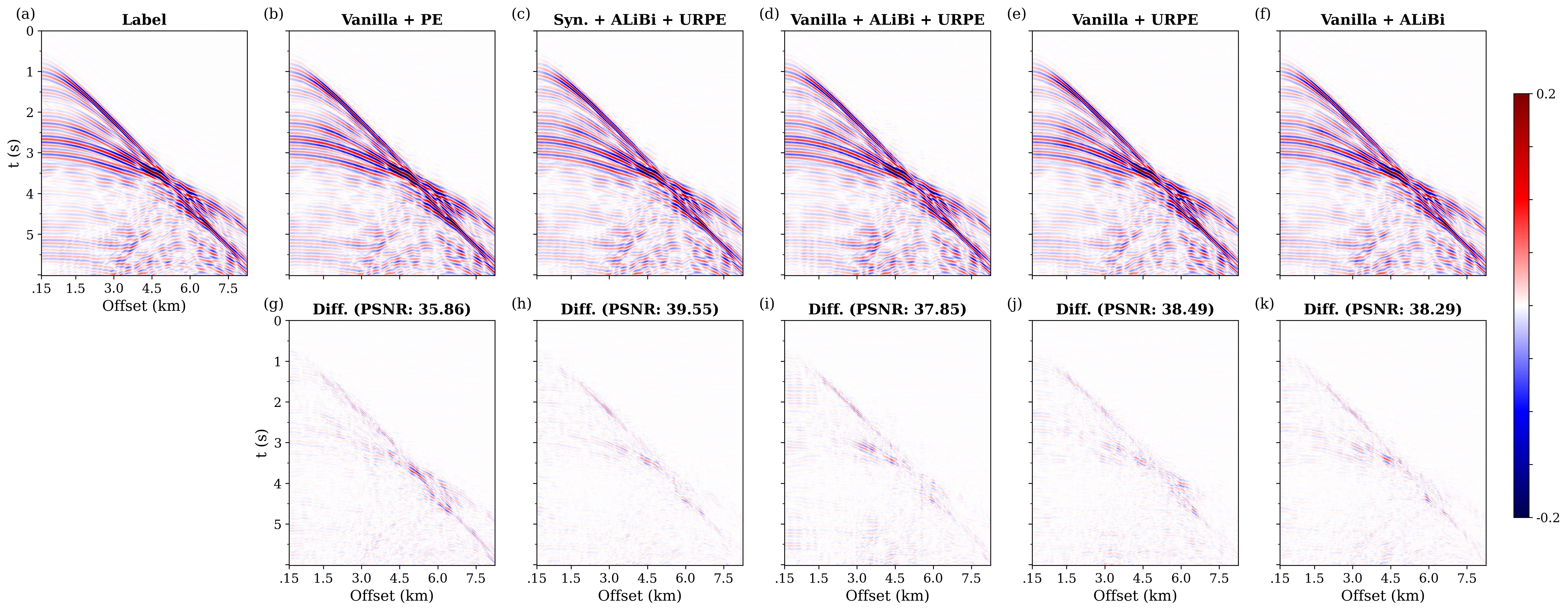}
    \caption{An example of dedirect task on the Marmousi data. The input to each NN model is the corresponding output in Figure 4(b--f). (a) The corresponding direct arrival removed label. The output of: (b) Vanilla + PE; (c) Syn. + ALiBi + URPE; (d) Vanilla + ALiBi + URPE; (e) Vanilla + URPE, and; (f) Vanilla + ALiBi. (g--k) The difference between (b--f) and (g), respectively.}
    \label{fig:fig5}
\end{figure}

Next, for the direct arrival removal (dedirect) task, we use a similar training setup as in the previous task. Referring to Table \ref{tab:tab2} that listed the results, again we observe that the Vanilla + PE model failed to generalize well to the test (Marmousi) set. For this task, the Syn. + ALiBi + URPE model achieved the highest metric in the test set. Consistent with the results in Table \ref{tab:tab2}, the direct arrival removal produced by Syn. + ALiBi + URPE model has the highest PSNR and less signal leakage amongst all (Figure \ref{fig:fig5}).

\begin{figure}[!h]
    \centering
    \includegraphics[width=1\textwidth]{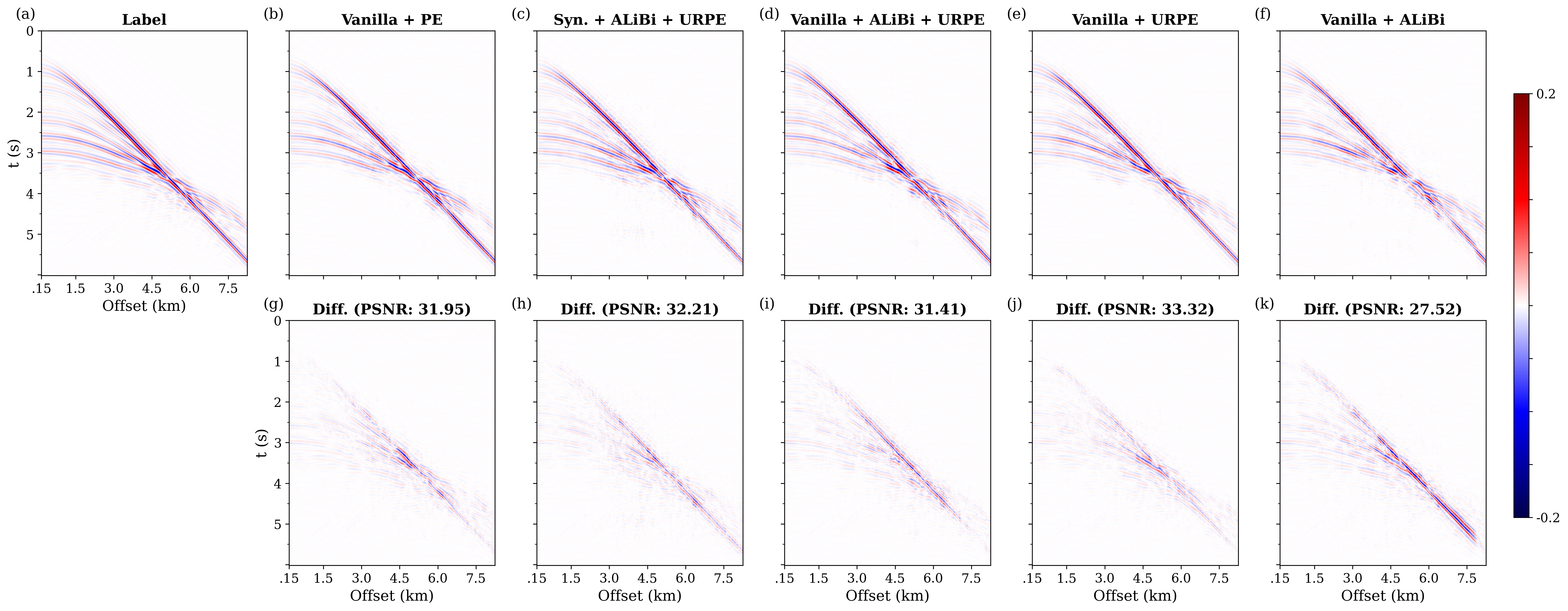}
    \caption{An example of demultiple task on the Marmousi data. The input to each NN model is the corresponding output in Figure 5(b--f). (a) The corresponding multiple-free label. The output of: (b) Vanilla + PE; (c) Syn. + ALiBi + URPE; (d) Vanilla + ALiBi + URPE; (e) Vanilla + URPE, and; (f) Vanilla + ALiBi. (g--k) The difference between (b--f) and (g), respectively.}
    \label{fig:fig6}
\end{figure}

We fine-tune the pretrained model for a multiple attenuation (demultiple) task as explained in Section \ref{sec:theory_3}. From Table \ref{tab:tab2}, we could observe that for this task, the Vanilla + URPE model achieved the highest PSNR in both the validation and test sets. Similar to the denoising task, the Syn. + ALiBi + URPE model achieved a PSNR of 33.26 dB, close to the Vanilla + PE model with a PSNR of 33.48 dB in second place. A sample result is shown in Figure \ref{fig:fig6}. While the Vanilla + URPE demultiple data have the highest PSNR, the Syn. + ALiBi + URPE and Vanilla + PE demultiple data offer competitive results.

\begin{figure}[!h]
    \centering
    \includegraphics[width=0.4\textwidth]{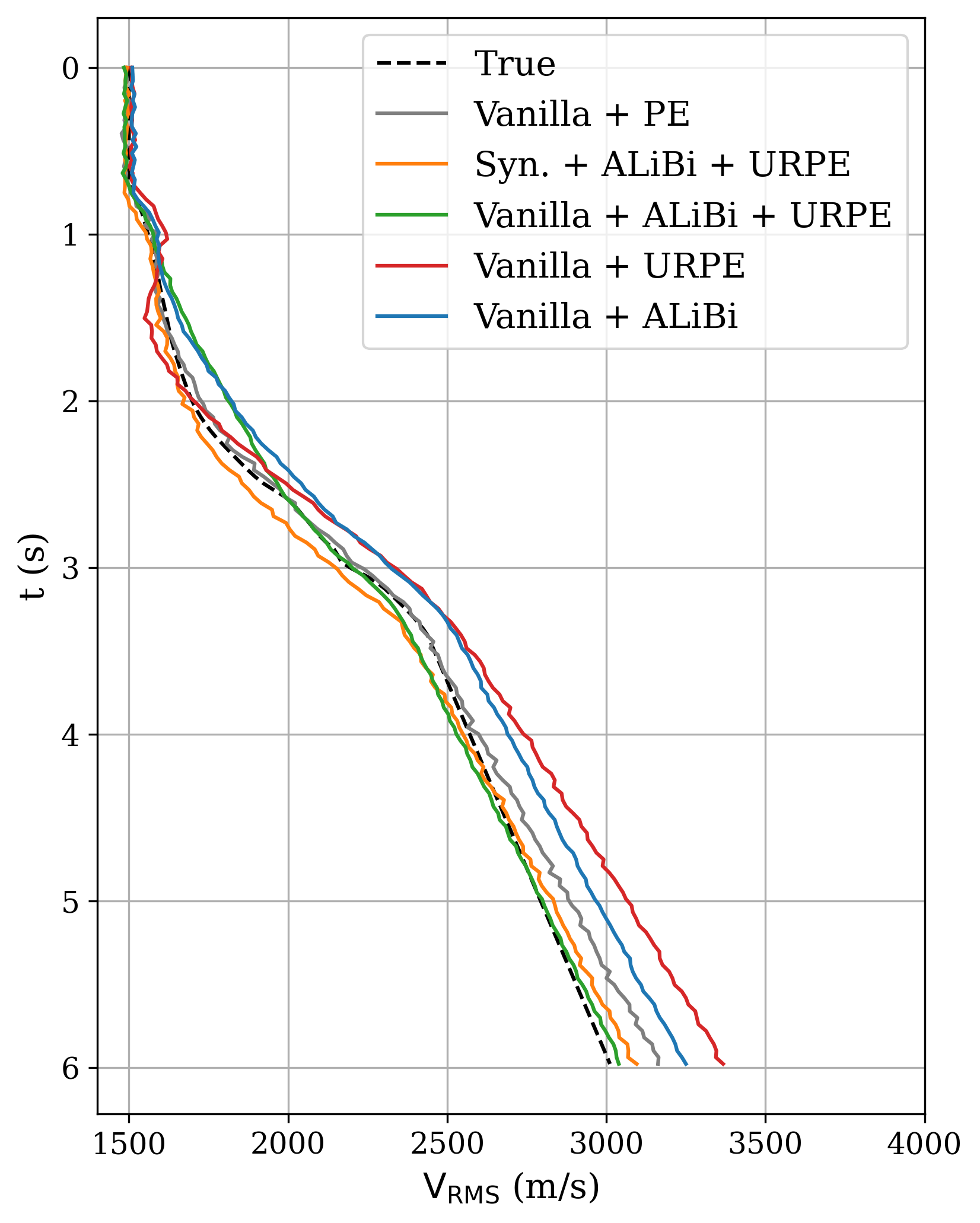}
    \caption{An example of the $V_{RMS}$ prediction task on the Marmousi data. The input to each NN model is the corresponding output in Figure 6(b--f). The black dashed line represents the true $V_{RMS}$ used as the label. The colored lines are the output of Vanilla + PE (grey), Syn. + ALiBi + URPE (orange), Vanilla + ALiBi + URPE (green), Vanilla + URPE (red), and Vanilla + ALiBi (blue).}
    \label{fig:fig7}
\end{figure}

\begin{figure}[!h]
    \centering
    \includegraphics[width=1\textwidth]{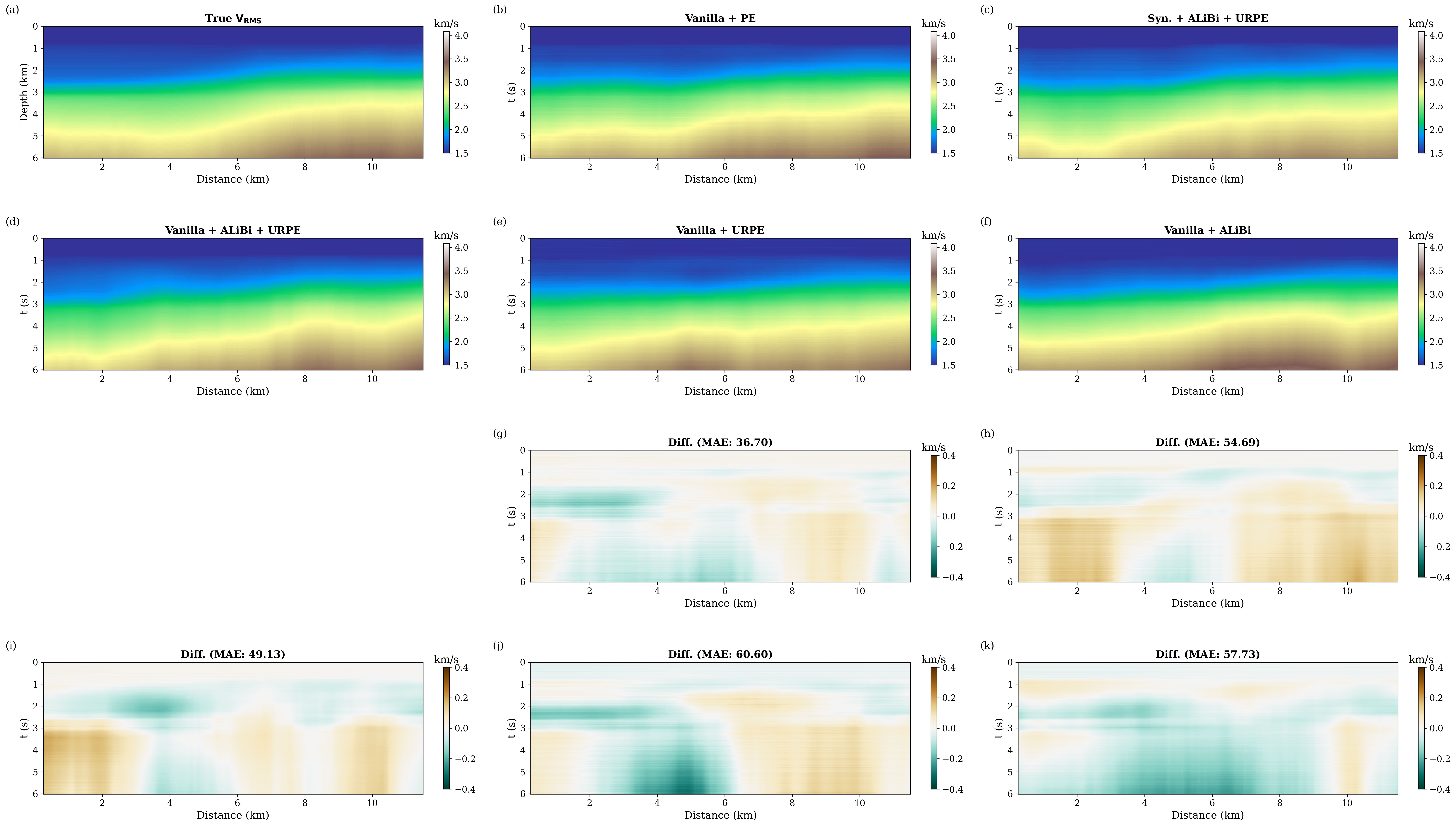}
    \caption{(a) True $V_{RMS}$ profile of the Marmousi model. Stacked section of the predicted $V_{RMS}$ from: (b) Vanilla + PE; (c) Syn. + ALiBi + URPE; (d) Vanilla + ALiBi + URPE; (e) Vanilla + URPE, and; (f) Vanilla + ALiBi. (g--k) The difference between (b--f) with (a), respectively.}
    \label{fig:fig8}
\end{figure}

The last task for this experiment, predicting the $V_{RMS}$, utilizes similar procedures as the previous tasks. However, for this task, we use an L1 loss function, which seems to work better for velocities. The Syn. + ALiBi + URPE achieved the lowest MAE (Mean Absolute Error) of 67.03 m/s in the test set (Table \ref{tab:tab2}). Though not performing equivalently with the Vanilla + PE model that achieved an MAE of 49.17 m/s in the test set, the Syn. + ALiBi + URPE is the second among the list. Figure \ref{fig:fig7} shows an example of the predictions from each NN model. The Vanilla + PE model fitted the labels almost perfectly at shallow depths but deviated slightly at deeper depths. On the other hand, the predicted $V_{RMS}$ from Syn. + ALiBi + URPE model is better at later times but worse up shallow. In Figure \ref{fig:fig8}, we show the 2D predicted $V_{RMS}$ profiles from all NN models, obtained through stacking and averaging over all samples. Based on the residuals shown in Figure \ref{fig:fig8}(g--k), we could infer that the results from Vanilla + PE, Syn. + ALiBi + URPE, and Vanilla + ALiBi + URPE are the best three, with Vanilla + PE predictions still having the lowest MAE.

\begin{figure}[!h]
    \centering
    \includegraphics[width=1\textwidth]{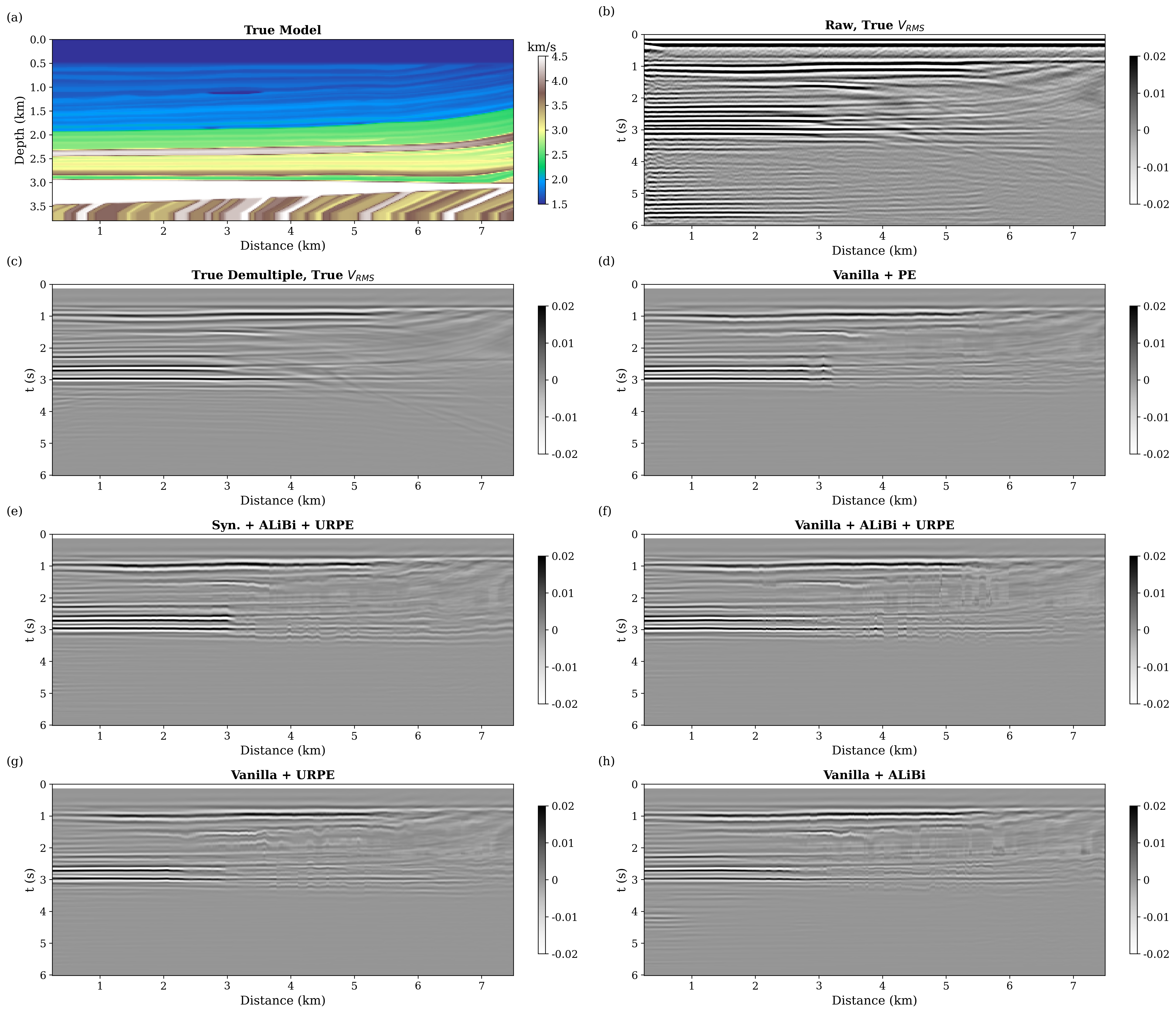}
    \caption{(a) True Marmousi velocity model. Stacked section of: (b) the raw Marmousi data, corrected with true $V_{RMS}$; (c) the true processed Marmousi data, corrected with true $V_{RMS}$; (d) the processed Marmousi data using Vanilla + PE, corrected with the corresponding predicted $V_{RMS}$; (e) the processed Marmousi data using Syn. + ALiBi + URPE, corrected with the corresponding predicted $V_{RMS}$; (f) the processed Marmousi data using Vanilla + ALiBi + URPE, corrected with the corresponding predicted $V_{RMS}$; (g) the processed Marmousi data using Vanilla + URPE, corrected with the corresponding predicted $V_{RMS}$, and; (h) the processed Marmousi data using Vanilla + ALiBi, corrected with the corresponding predicted $V_{RMS}$.}
    \label{fig:fig9}
\end{figure}

Additionally, we perform NMO correction and stacking on common midpoint gathers on the left portion of the Marmousi using the predicted $V_{RMS}$ to produce a post-stack section, shown in Figure \ref{fig:fig9}. The stacked sections of all NN models look satisfactory when compared to the multiple-free stacked section (Figure \ref{fig:fig9}(c)), and most of the unwanted signals (direct arrival \& surface-related multiples) and noise from the raw data (Figure \ref{fig:fig9}(b)) are removed. We also include the true Marmousi velocity model corresponding to the images in Figure \ref{fig:fig9}(a). With a closer look, the image produced by Vanilla + PE (Figure \ref{fig:fig9}(d)) and Syn. ALiBi + URPE (Figure \ref{fig:fig9}(e)) are cleaner than the others, especially at the right part of the image ($\geq$ 5 km).

Finally, we compare the total training time (pretraining and fine-tuning) of each model, listed in the last column of Table \ref{tab:tab2}. The Vanilla + PE model spent the longest time for training (32.1 hours) due to the slow convergence in the pretraining. Replacing the PE with ALiBi leads to a much shorter total training time (14.1 hours). Models with URPE need longer training time due to the additional cost of forming the Toeplitz matrix. However, this can be alleviated using the factored synthesizer, which makes a model converge faster in the pretraining, thus leading to the shortest total training time for the Syn. + ALiBi + URPE model (14.0 hours).

\subsection{North West Australia data example}
\label{sec:results_2}
We repeat the flow used for the Marmousi data to an offshore (marine) seismic data acquired in North West Australia by CGG. The data were acquired using a streamer containing 648 receivers with a spacing of 12.5 m, which produced 1,824 shot gathers with a spacing of 18.74 m and a sampling rate of 1 ms using an airgun-type source. We were also provided a well-log measurement located at 10.5 km along the acquisition line (Figure \ref{fig:fig13}), which we use as a reference to create 4,096 random velocity models. Using the original acquisition parameters, we simulate three shots per model using an elastic wave propagator \cite{richardson_alan_2023}, which results in 12,288 samples in our inventory. We resample both synthetic and field data to reduce the number of receivers to 324 and the number of time samples from 6,016 to 376. 

For this example, we will only show the results of the vanilla model and the fully modified model (synthesizer + ALiBi + URPE). We only used an equal number of synthetic and field shot gathers (1,824 samples each) in the pretraining, then applied the same augmentation techniques and training-test split as in Section \ref{sec:results_1}, resulting in a split of 52,452 and 13,140 samples for training and validation, respectively. The pretraining took 415 epochs and 15 hours for the Vanilla + PE model, and 108 epochs and 5.7 hours for the Syn. + ALiBi + URPE model. In the fine-tuning, we use the whole 12,288 samples, augment them as in Section \ref{sec:results_1}, and as a result obtain a split of 19,432 and 4,858 samples for training and validation, respectively. The training procedures and parameters are all similar to that explained in Section \ref{sec:results_1}.

\begin{figure}[!h]
    \centering
    \includegraphics[width=0.75\textwidth]{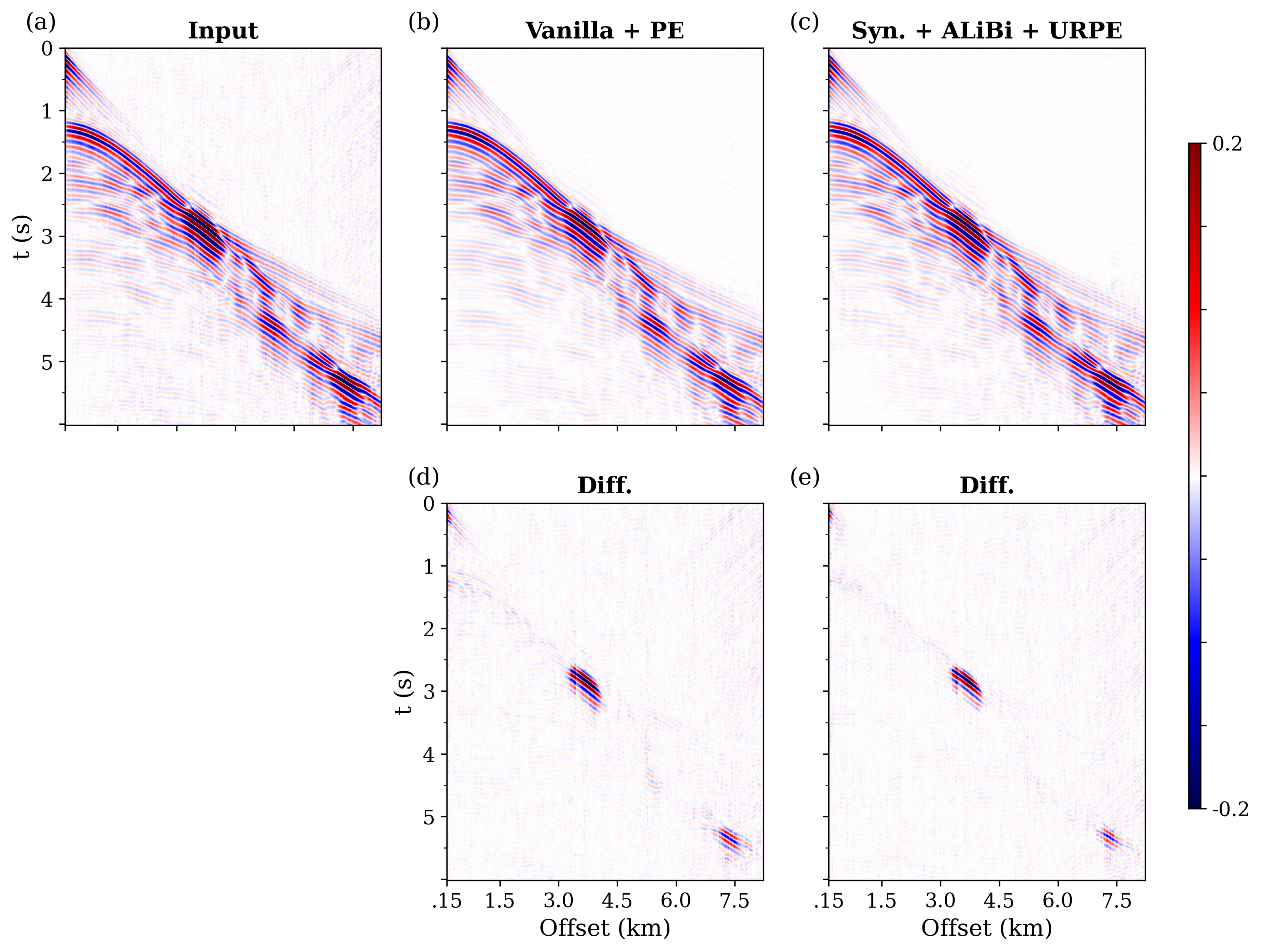}
    \caption{An example of denoising task on the field data. (a) Noisy input for each NN model. The output of: (b) Vanilla + PE, and; (c) Syn. + ALiBi + URPE. (d--e) The difference of (b--c) with (a), respectively.}
    \label{fig:fig10}
\end{figure}

\begin{figure}[!h]
    \centering
    \includegraphics[width=0.5\textwidth]{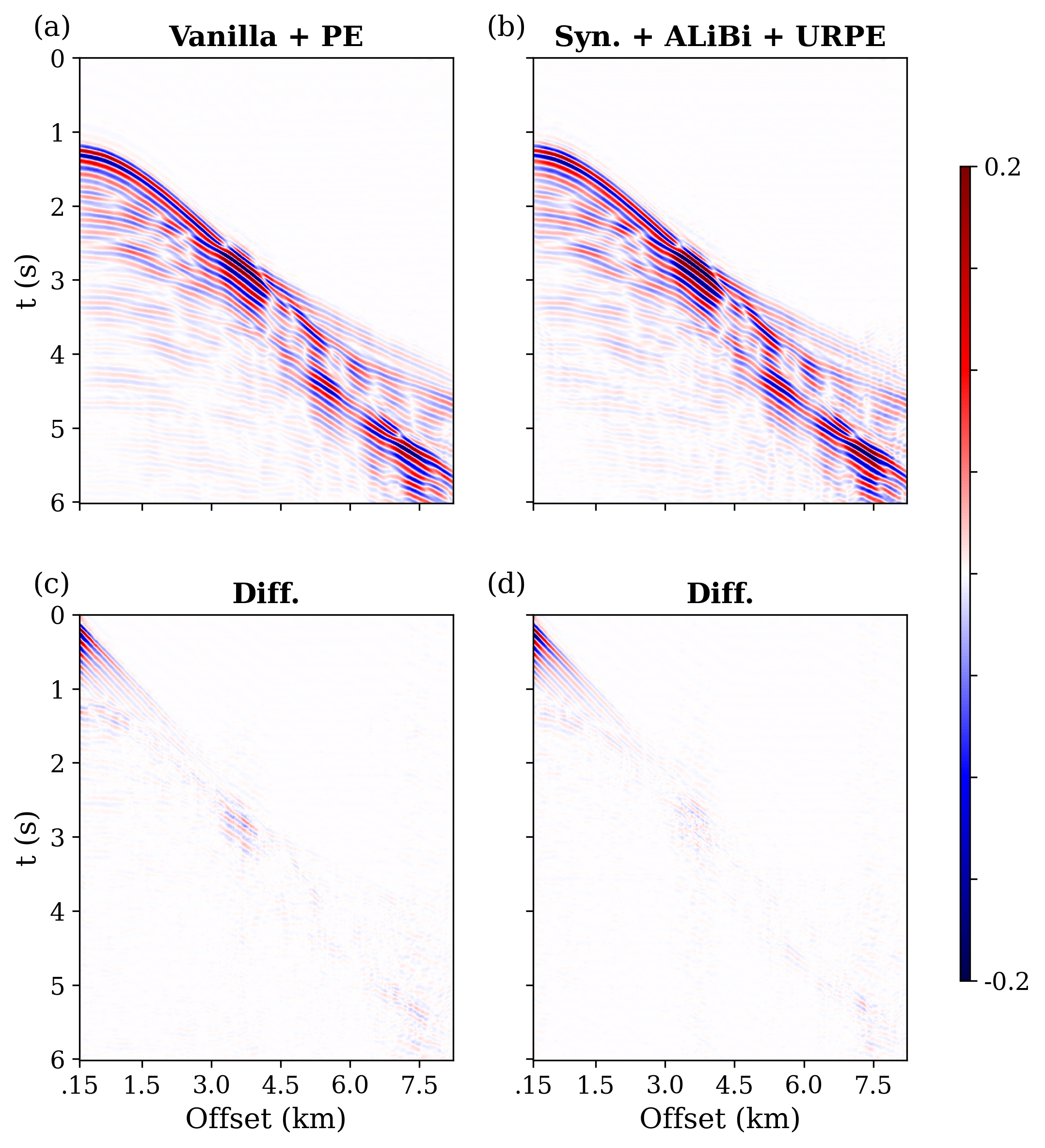}
    \caption{An example of dedirect task on the field data. The input to each NN model is the corresponding output in Figure 9(b--c). The output of: (b) Vanilla + PE, and; (c) Syn. + ALiBi + URPE. (c--d) The difference between (a--b) and Figure 9(b--c), respectively.}
    \label{fig:fig11}
\end{figure}

\begin{figure}[!h]
    \centering
    \includegraphics[width=0.5\textwidth]{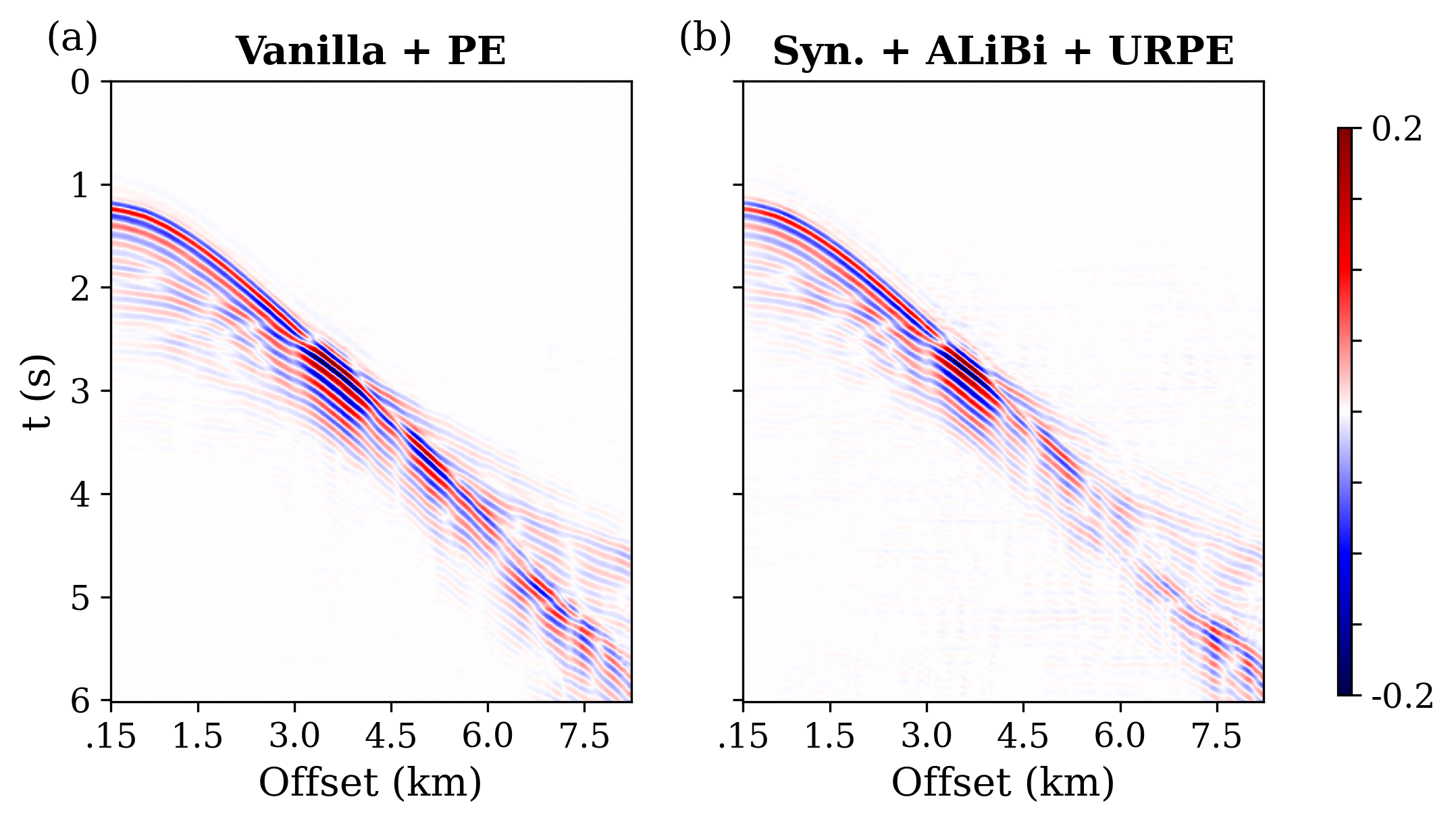}
    \caption{An example of demultiple task on the field data. The input to each NN model is the corresponding output in Figure 10(a--b). The output of: (b) Vanilla + PE, and; (c) Syn. + ALiBi + URPE.}
    \label{fig:fig12}
\end{figure}

\begin{figure}[!h]
    \centering
    \includegraphics[width=0.4\textwidth]{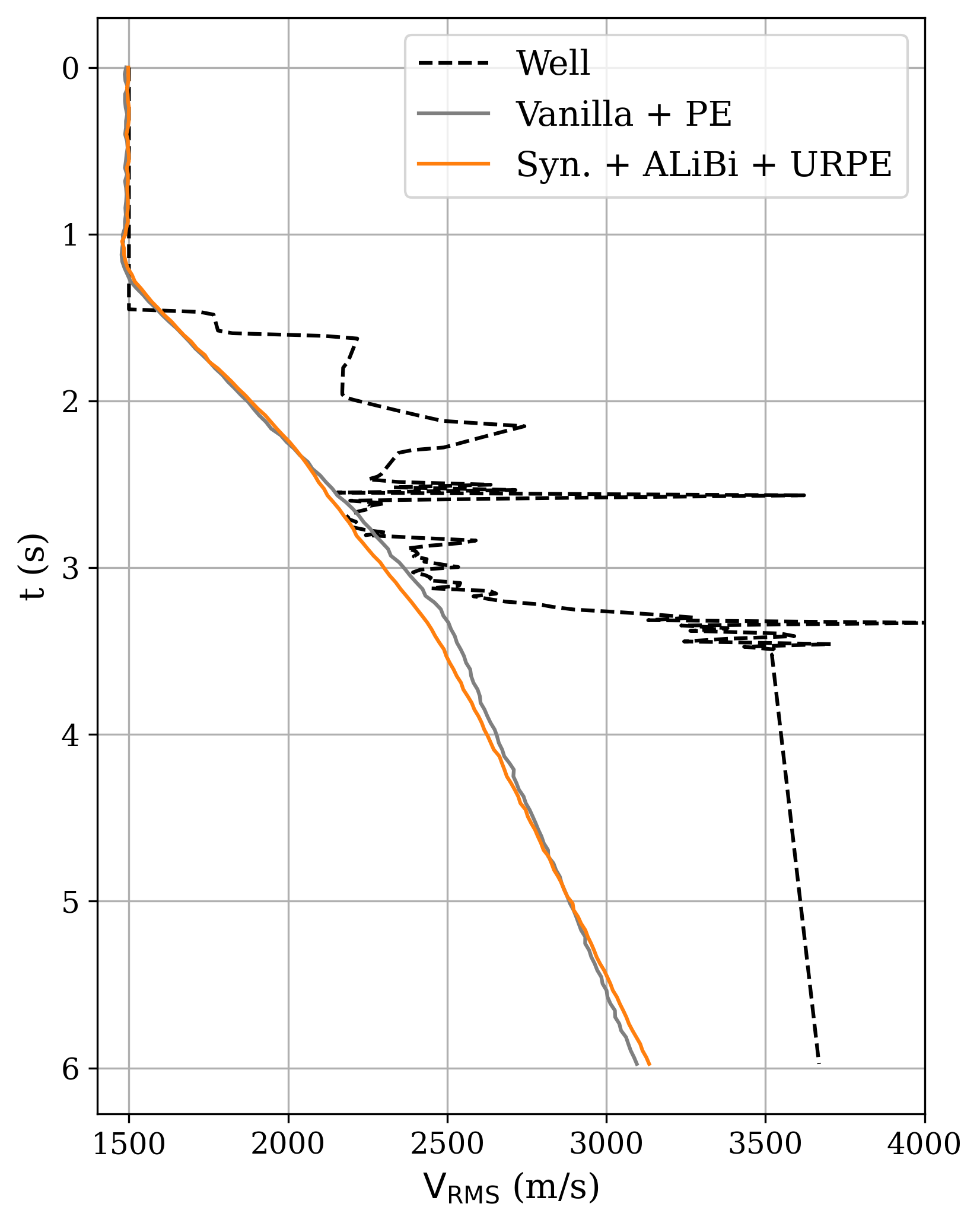}
    \caption{An example of the $V_{RMS}$ prediction task on the field data. The input to each NN model is the corresponding output in Figure 10(a--b). The black dashed line represents the well-log measurement used as a reference. The colored lines are the output of Vanilla + PE (grey) and Syn. + ALBi + URPE (orange).}
    \label{fig:fig13}
\end{figure}

\begin{figure}[!h]
    \centering
    \includegraphics[width=1\textwidth]{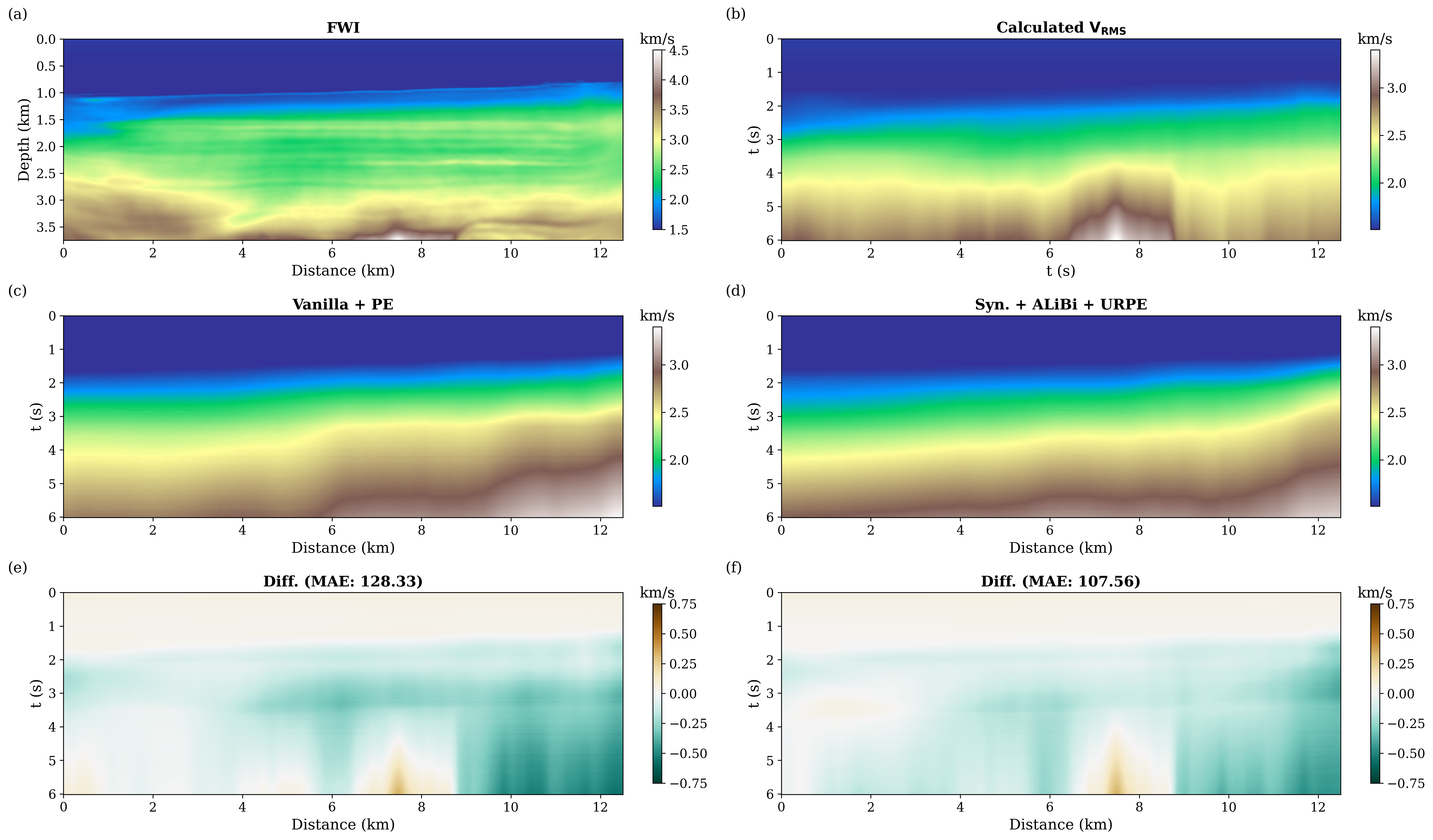}
    \caption{(a) FWI result of \cite{kalita2017efficient}. (b) $V_{RMS}$ profile calculated from (a). Stacked section of the predicted $V_{RMS}$ from: (c) Vanilla + PE, and; (d) Syn. + ALiBi + URPE. (e--f) The difference between (c--d) with (b), respectively.}
    \label{fig:fig14}
\end{figure}

\begin{figure}[!h]
    \centering
    \includegraphics[width=1\textwidth]{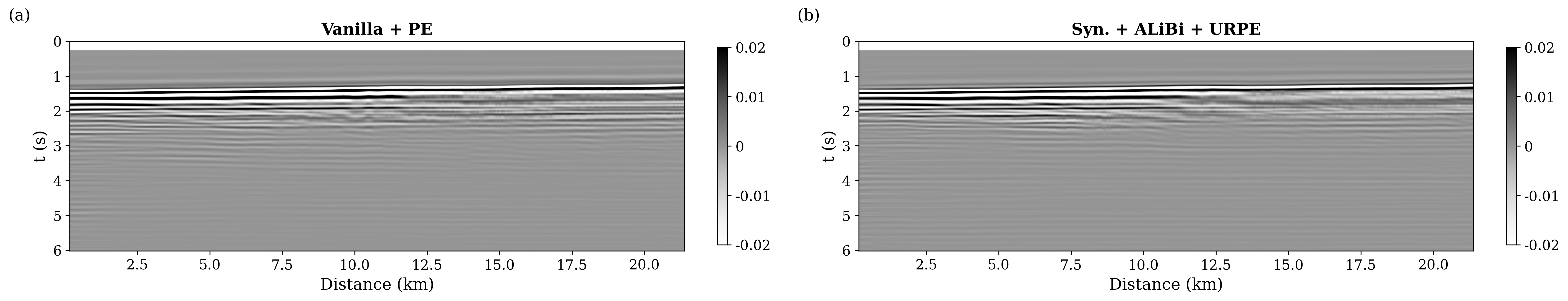}
    \caption{Stacked section of: (a) processed field data using Vanilla + PE, corrected with the corresponding predicted $V_{RMS}$, and; (b) processed field data using Syn. + ALiBi + URPE, corrected with the corresponding predicted $V_{RMS}$.}
    \label{fig:fig15}
\end{figure}

The Syn. + ALiBi + URPE denoised data contain less signal leakage than that of the Vanilla + PE model (Figure \ref{fig:fig10}), retaining the same feature we observed in Section \ref{sec:results_1}. Similarly, the Syn. + ALiBi + URPE model managed to remove the direct arrival with less signal leakage compared to the Vanilla + PE model (Figure \ref{fig:fig11}). In the demultiple task, however, the Vanilla + PE model produces stronger signals at later times compared to the Syn. + ALiBi + URPE (Figure \ref{fig:fig12}). For the $V_{RMS}$ prediction task, both models produced quite similar velocity profiles (Figure \ref{fig:fig13}). We performed the same stacking procedure to the predicted $V_{RMS}$ as in Section \ref{sec:results_1}, shown in Figure \ref{fig:fig14}(c--d), and compare them with $V_{RMS}$ profile calculated from full waveform inversion (FWI) result of \cite{kalita2017efficient} (Figures \ref{fig:fig14}(b) and \ref{fig:fig14}(a), respectively). The velocities are reasonably reproduced at the left region, though they have notable disagreements at the right region. Nevertheless, the Syn. + ALiBi + URPE predictions are slightly better than that of the Vanilla + PE (Figure \ref{fig:fig14}(e--f)). The post-stack sections from both models, shown in Figure \ref{fig:fig15}, are also comparable, with the image from Syn. + ALiBi + URPE is better resolved at the right part ($\geq$ 12.5 km).

\section{Discussion}
\label{sec:discussions}
Sections \ref{sec:introduction} - \ref{sec:results} scrutinized the two key aspects in this study, which are the sequential implementation of a pretrained model in seismic processing tasks and the proposed alternatives to improve StorSeismic. This section will answer the questions that led to these findings: 1). \textit{How significant is the pretraining step in the sequential implementation?} and 2). \textit{How sensitive are the parameters of the modified components (synthesizer and ALiBi)?} To quantitatively answer these questions, we perform ablation studies using the data generated from the Marmousi model (Section \ref{sec:results_1}).

\subsection{The importance of pretraining}
\label{sec:discussions_1}
To show the significance of the pretraining step, we redo all fine-tuning tasks using the same setup, but we use randomly-initialized models instead of the pretrained model as a starting point for each task. Table \ref{tab:tab3} shows the comparison of the results for the test set. 

\begin{table}[!h]
    \centering
    \begin{tabular}{c|c|c|c|c|c|c|c|c}
        \multirow{3}{*}{} & \multicolumn{2}{c|}{Denoising} & \multicolumn{2}{c|}{Dedirect} & \multicolumn{2}{c|}{Demultiple} & \multicolumn{2}{c}{$V_{RMS}$ prediction}\\
        & \multicolumn{2}{c|}{(PSNR $\uparrow$)} & \multicolumn{2}{c|}{(PSNR $\uparrow$)} & \multicolumn{2}{c|}{(PSNR $\uparrow$)} & \multicolumn{2}{c}{(MAE $\downarrow$)}\\
        \cline{2-9}
        & Valid. & Test & Valid. & Test & Valid. & Test & Valid. & Test \\
        \hline\hline
        Without pretraining & 36.47 & 35.92 & 40.64 & 39.28 & 39.82 & 30.71 & 68.49 & 91.33 \\
        With pretraining & \textbf{41.33} & \textbf{41.02} & \textbf{40.97} & \textbf{41.14} & \textbf{43.12} & \textbf{33.26} & \textbf{67.03} & \textbf{61.23} \\
    \end{tabular}
    \caption{Measured metrics on all fine-tuning tasks for Syn. + ALiBi + URPE model without pretraining and with pretraining. The numbers shown are PSNR for denoising, dedirect, and demultiple, and MAE for $V_{RMS}$ prediction, averaged over the whole corresponding set.}
    \label{tab:tab3}
\end{table}

The resulting metrics of all of the fine-tuning tasks of the pretrained model are better than that of the randomly-initialized models, especially in the $V_{RMS}$ prediction task where we observe a significant performance boost on the pretrained model. This proves that the pretraining step is essential for the model to store the features of the seismic data. Although the pretraining step is costly for the vanilla architecture, through this study, we proved that the pretraining cost could be reduced by using the proposed alternative model.

\subsection{The choice of parameter \textbf{\textit{k}} of the factorized synthesizer}
\label{sec:discussions_2}
\begin{figure}[!h] 
\centering
\subfigure[]{\includegraphics[height=4cm]{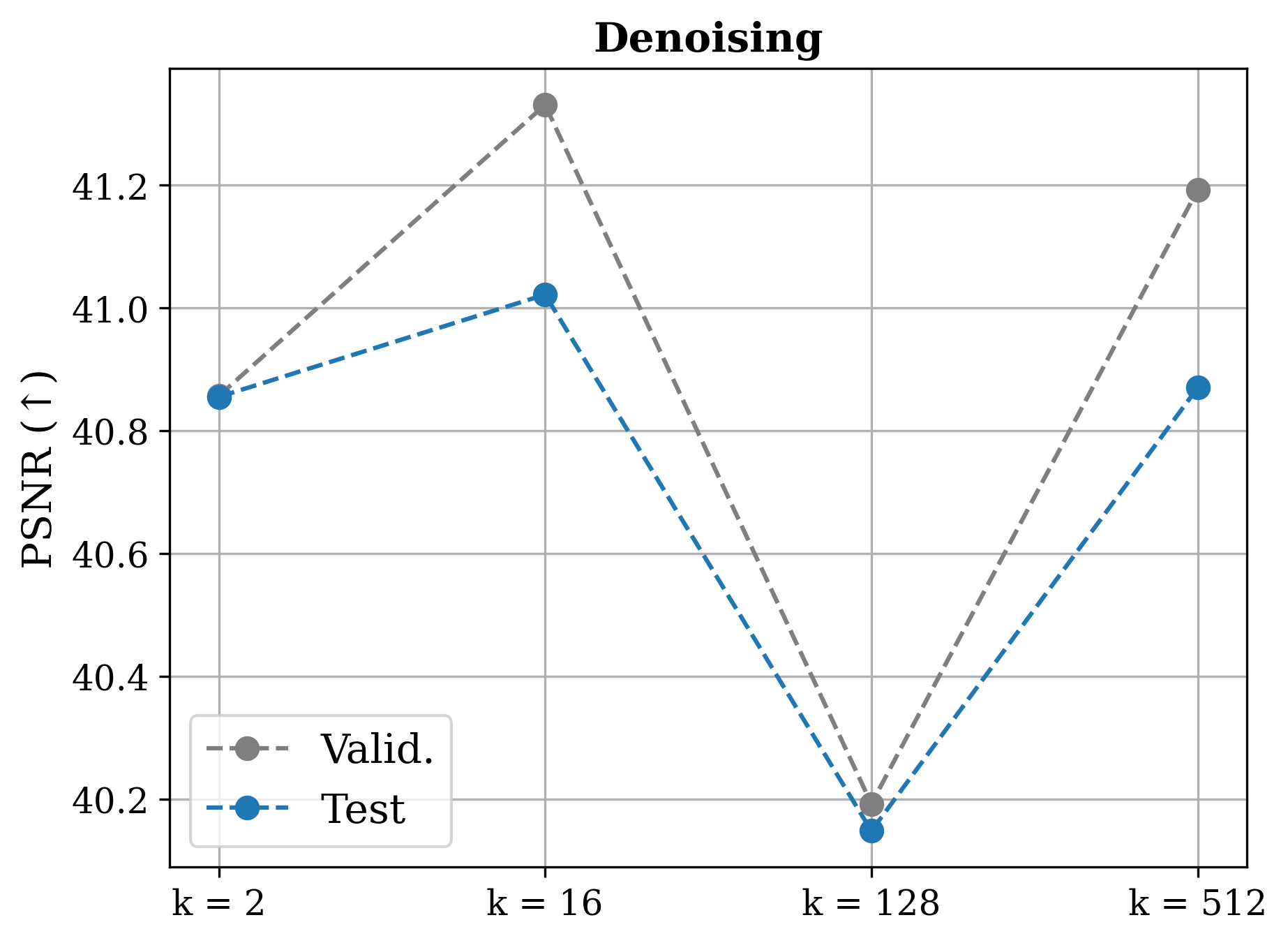} \label{fig:fig16a}}
\subfigure[]{\includegraphics[height=4cm]{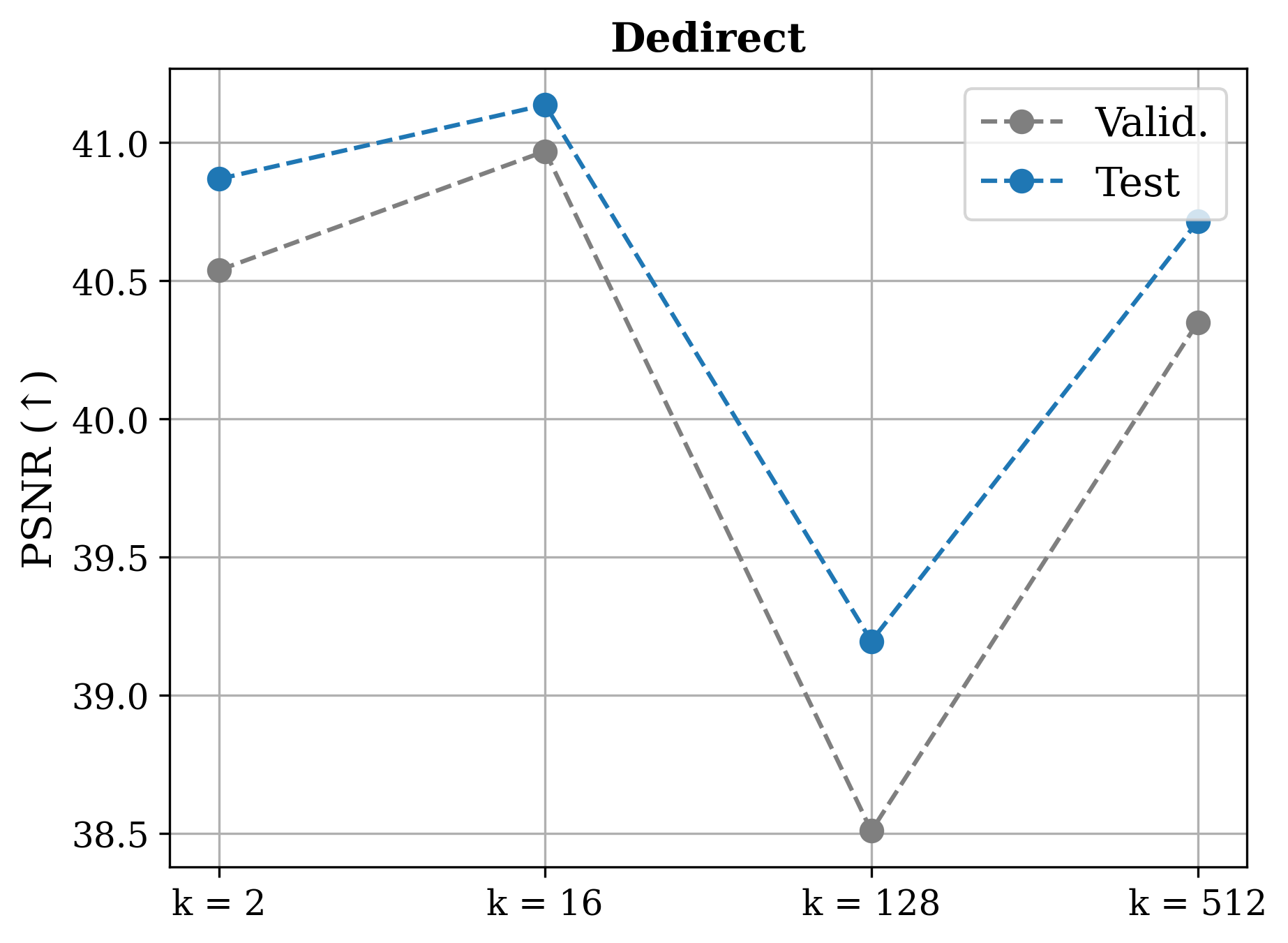} \label{fig:fig16b}}
\subfigure[]{\includegraphics[height=4cm]{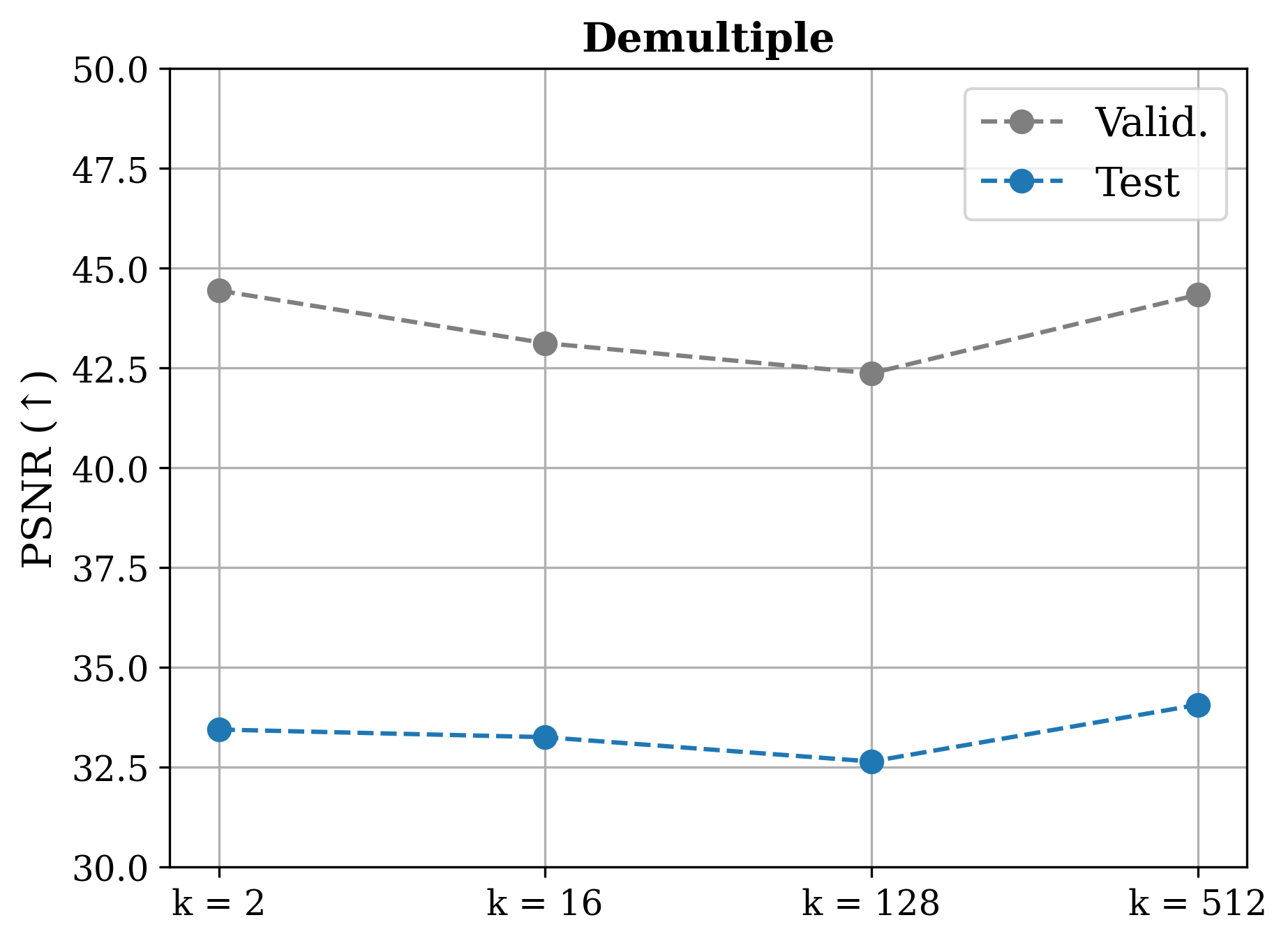} \label{fig:fig16c}}
\subfigure[]{\includegraphics[height=4cm]{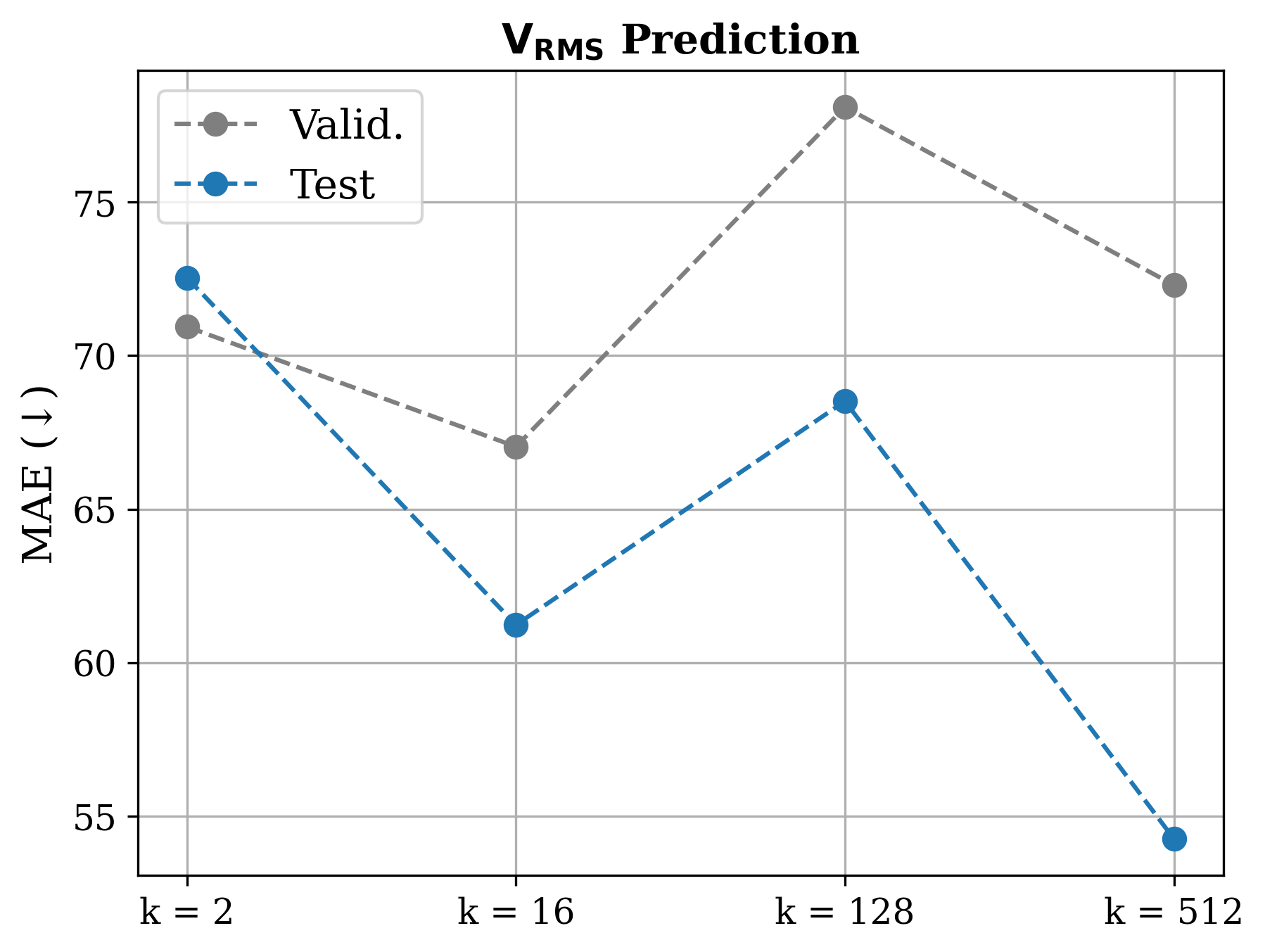} \label{fig:fig16d}}
\subfigure[]{\includegraphics[height=4cm]{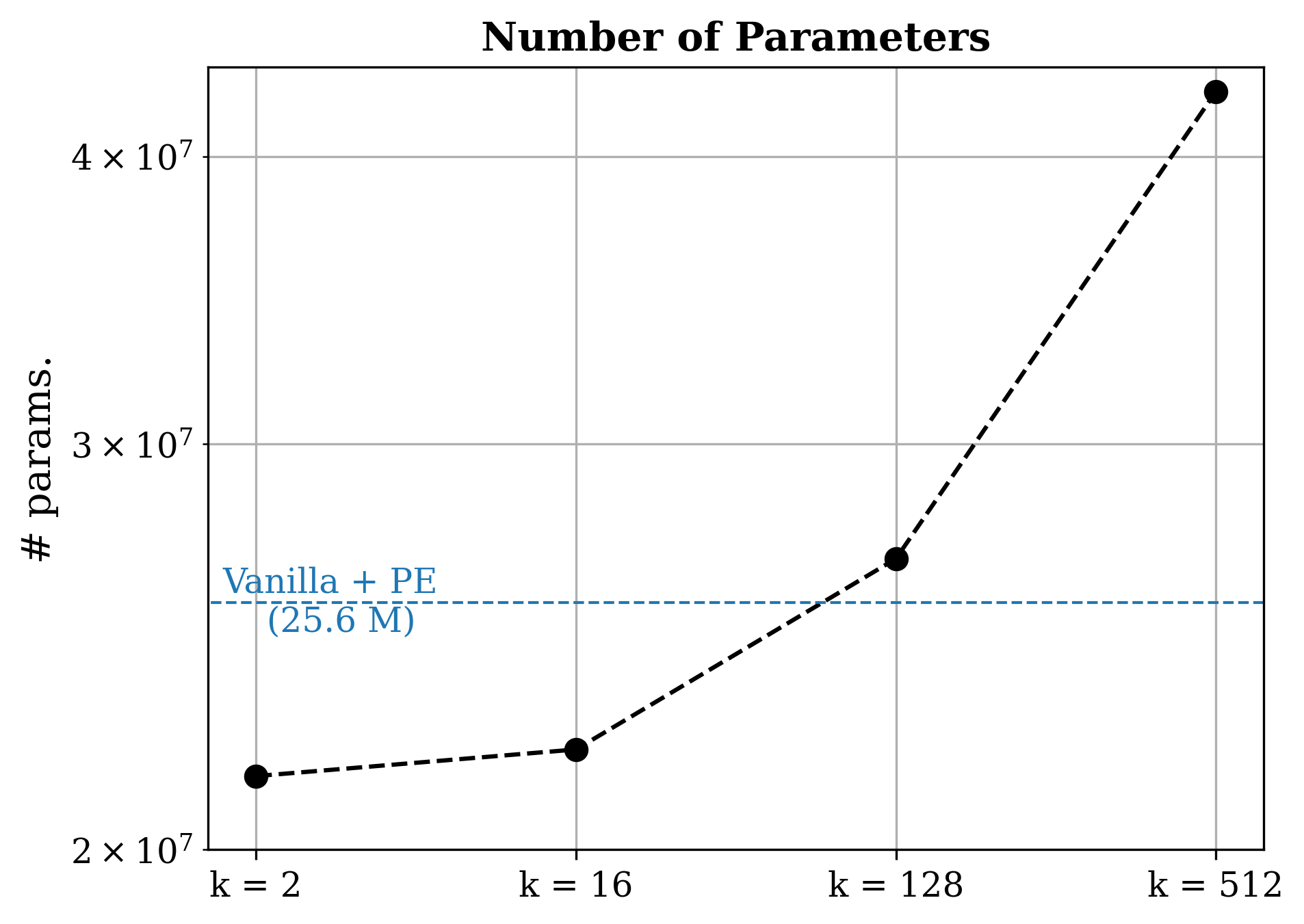} \label{fig:fig16e}} 
\caption{Measured metrics (y-axis) on the validation set (grey lines) and the test set (blue lines) for each model with different values of k (x-axis) on: the (a) denoising task; (b) the dedirect task; (c) the demultiple task, and; (d) the $V_{RMS}$ prediction task. (e) Number of parameters (y-axis) for each of the tested $k$ values (x-axis), with the blue dashed lines marking the number of parameters of the Vanilla + PE model for reference.} 
\label{fig:fig16} 
\end{figure}

As mentioned in Section \ref{sec:theory_2}, the factorized synthesizer has a tunable parameter $k$, which controls the rank of the attention matrix. With $k = N$, the attention matrix can become fully-ranked analogous to the weights of the query (Q) and the key (K) matrices in the vanilla architecture. Here, we  test different values of $k$ to analyze its sensitivity to the results, summarized in Figure \ref{fig:fig16}.

Directly looking at the number of parameters (Figure \ref{fig:fig16e}), we observe that by using $k \geq 128$, the number of parameters of the fully modified model superseded that of the vanilla model, which is attributed to the dependency of the synthesizer to the number of attention heads ($A$, refer to Section \ref{sec:theory_2}). Hence, we observe in Figure \ref{fig:fig16}(a--d) that there are "false" improvements in the fine-tuning tasks when using $k = 512$, likely due to the over-parameterization of the model.

Comparing three values of $k = 2$, $k = 16$, and $k = 128$, we obtain the best result across all tasks when using $k = 16$ (Figure \ref{fig:fig16}(a--d)). We found that this is the optimal value to balance between under-parameterization of the model and preventing overfitting of the model \cite{tay2021synthesizer}. Therefore, we use $k = 16$ for all models that utilize the factored synthesizer.

\subsection{Learnable vs. fixed slopes of the ALiBi}
\label{sec:discussions_3}
The ALiBi was initially proposed by \cite{press2021train} with fixed, non-learned slopes. They also mentioned an alternative in which the slopes are learnable. We test these two variations of ALiBi in our work, with the results summarized in Table \ref{tab:tab4}.

\begin{table}[!h]
    \centering
    ~\clap{
    \begin{tabular}{c|c|c|c|c|c|c|c|c}
        \multirow{3}{*}{} & \multicolumn{2}{c|}{Denoising} & \multicolumn{2}{c|}{Dedirect} & \multicolumn{2}{c|}{Demultiple} & \multicolumn{2}{c}{$V_{RMS}$ prediction}\\
        & \multicolumn{2}{c|}{(PSNR $\uparrow$)} & \multicolumn{2}{c|}{(PSNR $\uparrow$)} & \multicolumn{2}{c|}{(PSNR $\uparrow$)} & \multicolumn{2}{c}{(MAE $\downarrow$)}\\
        \cline{2-9}
        & Valid. & Test & Valid. & Test & Valid. & Test & Valid. & Test \\
        \hline\hline
        Syn. + ALiBi (fixed) + URPE & 40.72 & 40.63 & 40.58 & 41.02 & 42.33 & 31.83 & 70.52 & 73.11 \\
        Syn. + ALiBi (learnable) + URPE & \textbf{41.33} & \textbf{41.02} & \textbf{40.97} & \textbf{41.14} & \textbf{43.12} & \textbf{33.26} & \textbf{67.03} & \textbf{61.23} \\
    \end{tabular}
    }
    \caption{Measured metrics on all fine-tuning tasks for Syn. + ALiBi + URPE model with fixed slopes and with learnable slopes. The numbers shown are PSNR for denoising, dedirect, and demultiple, and MAE for $V_{RMS}$ prediction, averaged over the whole corresponding set.}
    \label{tab:tab4}
\end{table}

We clearly observe that the model with learnable slopes of ALiBi performs better than the fixed slopes in all of the fine-tuning tasks. The learnable slopes give more expressiveness to the network, making it adapt robustly to various tasks. Moreover, the additional parameters for the learnable slopes are only linearly proportional to the number of attention heads ($A$), which is negligible compared to the overall cost.

\section{Conclusions}
\label{sec:conclusions}
Large language models, represented mainly by the Transformer architecture, have gained a lot of attention due to their success across many NLP tasks. Inspired by this concept, we previously introduced StorSeismic, a Transformer-based architecture curated for seismic processing tasks, and showed the performance of the vanilla architecture on various seismic processing tasks through its pretraining and fine-tuning steps. In this study, we demonstrated the utilization of the network in a sequential seismic processing workflow, which takes in raw seismic data as input and performs denoising, direct arrival removal, multiple attenuation, and $V_{RMS}$ estimation, all through fine-tuning a pretrained StorSeismic model. Although we only show the application on the marine seismic acquisition on Marmousi and field data, adaptation to other types of seismic data is attainable through adjustments of the seismic processing steps that are appropriate for the data. 

We demonstrated how to improve the StorSeismic model in efficiency and effectiveness. Two modifications to the vanilla StorSeismic model were proposed: the low-rank attention mechanism, \textit{factored synthesizer}, to replace the vanilla self-attention, and; RPEs, \textit{ALiBi} and \textit{URPE}, to replace the sinusoidal PE. We showed that utilizing these modifications leads to faster convergence in the pretraining, less number of parameters, and competitive results on the fine-tuning tasks compared to the vanilla model. 

As there are many other alternatives to the vanilla self-attention and PE, other than those tested in this study, the authors suggest further investigation to explore the impact of these variations on the performance of the model. Additionally, modifications to other components of the proposed framework (e.g., modifying the training scheme) could also potentially lead to improvements.

\section*{Acknowledgment}
This publication is based on work supported by the King Abdullah University of Science and Technology
(KAUST). The authors thank the DeepWave sponsors for supporting this research. The authors also thank CGG for their permission to access the field data.

\bibliographystyle{unsrt}  
\bibliography{references}  

\end{document}